\documentclass[pre,aps,10pt,superscriptaddress,twocolumn,floatfix]{revtex4-1}

\usepackage[nice]{nicefrac}
\usepackage{graphicx,color}
\usepackage{amsmath,amssymb,bm,bbm}
\usepackage{amsmath}
\usepackage{braket}
\usepackage{bbold}
\usepackage{float}

\usepackage[plainpages=false,pdfpagelabels,colorlinks=true,linkcolor=red,urlcolor=blue,citecolor=blue,pdftitle={},pdfauthor={},pdfdisplaydoctitle=true,pdfduplex=DuplexFlipLongEdge]{hyperref}

\definecolor{darkred}{rgb}{0.90,0.2,0.2}
\definecolor{darkgreen}{rgb}{0,0.60,.2}
\definecolor{darkblue}{rgb}{0.1,0.3,1}
\definecolor{grey}{cmyk}{0,0,0,0.25}
\definecolor{orange}{cmyk}{0,0.6,0.8,0}

\begin{document}

\title{Eigenstate transitions and critical transport in 3D Aubry-André model}

\author{Pragna Das}
\affiliation{Department of Theoretical Physics, J. Stefan Institute, SI-1000 Ljubljana, Slovenia}
\author{Lev Vidmar}
\affiliation{Department of Theoretical Physics, J. Stefan Institute, SI-1000 Ljubljana, Slovenia}
\affiliation{Department of Physics, Faculty of Mathematics and Physics, University of Ljubljana, SI-1000 Ljubljana, Slovenia\looseness=-1}
\author{Miroslav Hopjan}
\affiliation{Department of Theoretical Physics, J. Stefan Institute, SI-1000 Ljubljana, Slovenia}
\affiliation{Institute of Theoretical Physics, Wroclaw University of Science and Technology, 50-370 Wrocław, Poland}

\begin{abstract}
We numerically study spectral and wave-function properties, as well as time evolution, in finite-size commensurate approximations of the 3D Aubry-André model. We explicitly show, through scaling analysis, the existence of two critical points.  One of the critical points is a localization-delocalization transition in real-space, similar to the critical point in the 3D Anderson model. The other is a localization-delocalization transition in a dual momentum space, which does not have any analog in the 3D Anderson model. We estimate their values from spectral measures, such as the gap ratio and the spectral form factor, as well as from wave-function measures, such as the inverse participation ratio and the participation entropy. Additionally, we demonstrate that both critical points exhibit hallmarks of critical points between quantum ergodicity and localization. First, the eigenstates and the eigenspectrum at the critical point fulfill the mutual universal relation between the eigenspectrum compressibility $\chi$ and the eigenstate fractal dimension $d^{(1)}$ [S. Jiricek et al., \href{https://doi.org/10.1073/pnas.2518027123}{Proc.~Nat.~Acad.~Sci.~\textbf{123}, e2518027123  (2026)}]. Second, the dynamics of the wave packet at the critical point is governed by the critical dynamical exponent $z$, characterized by the spectral fractal dimension and the dimensionality of the lattice [M. Hopjan and L. Vidmar, \href{https://doi.org/10.1103/PhysRevLett.131.060404}{Phys.~Rev.~Lett.~\textbf{135}, 060401  (2025)}]. Thus, similarly to the 3D Anderson model, the critical dynamical exponent is sub-diffusive with the value of $z=3$.
\end{abstract}

\maketitle

\section{Introduction}

The study of localization phenomena in quantum systems with disorder has been a cornerstone of condensed matter physics since the seminal work of Anderson$~$\cite{anderson1958absence} and the subsequent studies$~$\cite{edwards_thouless_72,Thouless74,Wegner76}. A milestone in understanding localization was the scaling theory of localization introduced in Ref.$~$\cite{Abrahams79}, where it was shown that delocalization can occur only in three dimensions (3D) and higher. Since then, the delocalization to localization transitions have been the subject of many theoretical and numerical studies$~$\cite{Wegner79,Altshuler80,MacKinnon81,MacKinnon83,Lee85,Altshuler88,Shklovskii93,kramer_mackinnon_93,Brandes96,Chalker96,Ohtsuki97,Slevin97b,Kawarabayashi98,Huckestein97,Huckestein98,Huckestein99,slevin_ohtsuki_99,Schubert05,markos_06,Rodriguez09,Rodriguez10,Ueoka14,Slevin_2014,Slevin16,Tarquini17,slevin_ohtsuki_18,Sierant2020,Zhao20,hopjan_orso_21,Herbrych21,suntajs_prosen_21,Horvath22,suntajs_prosen_23,hopjan2023,hopjan2023scale,Jiricek24,hopjan2024survival,hopjan2025critical,Altshuler25,jiricek2025universal}. It turned out that the delocalization-localization transition caused by disorder is a subclass of the broader class of transitions$~$\cite{jiricek2025universal}, where level statistics change from ergodic to Poisson and where the former is described by Random Matrix Theory (RMT)$~$\cite{Mirlin96, Mirlin00,Evers00,EversMirlin2008,fyodorov2009anderson,Bogomolny2011,jiricek2025universal}, such as the Gaussian orthonormal ensemble (GOE) or the Gaussian unitary ensemble (GUE) for real and complex Hamiltonians, respectively.

\begin{figure}[t!]
  \includegraphics[width=0.42\textwidth]{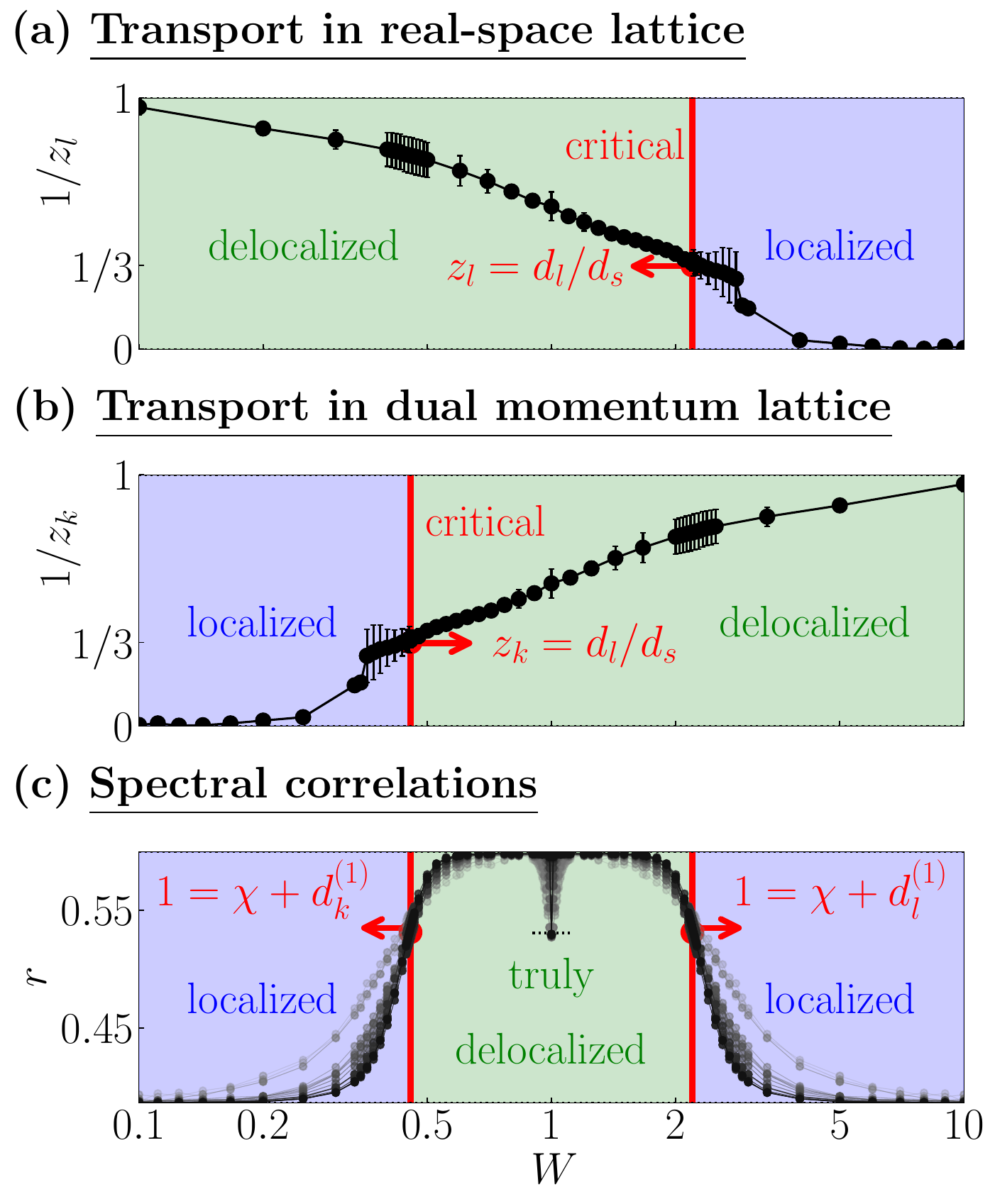}
  \caption{(a) and (b) The phases of the 3D Aubry-André models as seen by the transport in the real-space lattice, $l$, and in the dual momentum lattice, $k$, respectively. At criticality, the dynamical exponent $z$ fulfills prediction $z=d_l/d_s$ from Ref.$~$\cite{hopjan2025critical} with $d_l$ and $d_s$ being the lattice and spectral fractal dimension, respectively. (c) The phases of the same model as seen by spectral statistics, here, by the gap ratio $r$ of neighboring levels. At criticality, the universal relation $1=\chi+d^{(1)}$ from Ref.$~$\cite{jiricek2025universal}, between the eigenspectrum compressibility $\chi$ and the eigenstate fractal dimension $d^{(1)}$, is fulfiled. }
  \label{fig:intro}
\end{figure}

That being said, localization can also be induced by a different mechanism. A prominent example of a localized system without disorder is the quasiperiodic potential present in the celebrated one-dimensional (1D) Aubry-André (AA) model$~$\cite{Aubry80}. Here, the duality of the model between real and momentum space, due to quasiperiodicity, is the mechanism behind the transition$~$\cite{Aubry80}. The transition can be seen as a delocalization-to-localization transition in real space or, alternatively, as a localization-to-delocalization transition in momentum space. At the transition point, the model becomes dual and has peculiar properties such as a fractal eigenspectrum$~$\cite{harper1955single,hofstadter1976energy,Aubry80}.  The quasiperiodic Aubry-André model has been the subject of many theoretical and numerical studies$~$\cite{Suslov82,Thouless82,Kohmoto83,Chao86, Kohmoto87,Siebesma1987,Hiramoto89,Geisel_91,Ketzmerick_92,Hiramoto92, Ketzmerick_97, Macia2014, Li16, Dominguez-Castro_2019,hopjan_orso_21,wu2021,Zhang22,hopjan2023,hopjan2025critical} and was observed experimentally using cold atoms$~$\cite{Roati08,Luschen18} and photonic lattices$~$\cite{Lahini:PRL2009}.

Remarkably, see Fig.$~$\ref{fig:intro}, the features of the two kinds of localization phenomena are blended in the 3D Aubry-André model$~$\cite{devakul2017anderson,Sutradhar19,Luo22,Ticea24}. On the one hand, the study of transport properties reveals delocalization-to-localization and localization-to-delocalization transitions when considering transport in real space and momentum space, respectively; see Figs.$~$\ref{fig:intro}(a) and$~$\ref{fig:intro}(b). However, unlike in the 1D Aubry-André model, the transition points do not emerge at the selfdual point $W_{\rm SD}=1$, but rather at $W_c^{}$ and $1/W_c^{}$, with $W_c > W_{\rm SD}$. On the other hand, the study of spectral properties reveals that the transition points are of the ergodic-to-Poisson type and separate the truly delocalized ergodic phase from phases localized in at least one non-trivial basis; see Fig.$~$\ref{fig:intro}(c). We observe that the system flows towards the Poissonian level statistic with increasing system size for $W<1/W_c^{}$ and $W>W_c^{}$, while the system flows towards the predictions of the RMT (GUE or GOE) in a truly delocalized ergodic phase, $1/W_c^{}<W<W_c^{}$, with the crossing of $r$ around the points $W_c^{}$ and $1/W_c^{}$.

The main part of the present study is devoted to the accurate estimation of the value of the critical point $W_c^{}$ from the series of approximations that retain the dual properties of the infinite lattice. We note that this implementation is complementary to the implementations of Refs.$~$\cite{devakul2017anderson,Sutradhar19,Luo22}, where open boundary conditions were considered and, consequently, the explicit dual properties were lost. However, the estimates of the critical point $W_c$ and the correlation length exponent $\nu$ in Refs.$~$\cite{devakul2017anderson,Sutradhar19,Luo22} are close to our estimate here. 

To estimate the transition point in our implementation, we employ several diagnostics. We study spectral measures such as the gap ratio$~$\cite{oganesyan_huse_07,atas_bogomolny_13}, the spectral form factor with the associated Thouless time$~$\cite{suntajs_bonca_20a,suntajs_prosen_21,suntajs_vidmar_22, hopjan2023scale,Kliczkowski24,Swietek25,Swietek26,andreoni2026}, and the spectral compressibility$~$\cite{Chalker96, jiricek2025universal}. In order to determine the localization properties, we further study wave-function-based measures such as the inverse participation ratio$~$\cite{Mirlin00,Evers00} (IPR) and the participation entropy$~$\cite{mace2019multifractal,suntajs_prosen_21,suntajs2024}. By the finite-size scaling analysis, we show that the spectral measures and the wave-function-based measures point towards the same value of $W_c^{}$. 

With the accurate estimate at hand, we study the properties of the system at the estimated critical point. We first show that the universal relation between the spectral and wave-function properties at criticality$~$\cite{Chalker96,Bogomolny2011,jiricek2025universal} is fulfilled. Then, we turn to the transport properties at the critical point. We show that the spreading of the wave packet becomes governed by the critical dynamical exponent$~$\cite{hopjan2025critical} in the basis of localization. The fulfillment of the critical properties reinforces our estimates of the critical points.

The paper is organized as follows. We review and provide additional details about the model and the dual transform in Secs.$~$\ref{infinite system}-\ref{3D case transform twist}. In Secs.$~$\ref{3D case duality spectra}-\ref{3D case eigenstates}, we demonstrate the symmetries of the spectra and eigenstates. In Sec.$~$\ref{3D case edges}, we discuss the effects at spectral edges.  We then determine the transition points in Sec.$~$\ref{3D case}. In Sec.$~$\ref{3D case scaling}, we perform the finite-size scaling of the gap ratio and the inverse participation ratio. We determine the transition point value, the correlation length exponent, and the fractal dimension of the critical eigenstates. In Sec.$~$\ref{3D case spectral form factor} and Sec.$~$\ref{3D case filter}, we inspect the short-range and long-range spectral correlations, respectively, and compare them to theoretical predictions. Finally, in Sec.$~$\ref{3D case universal}, we demonstrate that the universal relation between spectral and wave-function properties$~$\cite{jiricek2025universal} is fulfilled at the estimated transition point. The dynamical properties are studied in Sec.$~$\ref{3D case time}. In Sec.$~$\ref{3D case time overlap} we demonstrate the symmetries of the transport. We study the scaling of the mean square displacement in Sec.$~$\ref{3D case time scaling} and its steady-state value in Sec.$~$\ref{3D case time steady state}. In Sec.$~$\ref{3D case time dynamical exponent}, we extract the dynamical exponent and demonstrate that the estimated transition point fulfills the prediction of the critical dynamical exponent$~$\cite{hopjan2025critical}. We conclude our study in Sec.$~$\ref{conclusions}.

\section{Model and symmetries} \label{model}

\subsection{Infinite-size system}
\label{infinite system}

Our goal is to determine the phases of the infinite-size cubic-lattice system where the kinetic and potential energies of a particle are described by the Hamiltonian 
\begin{equation}
    H_{\rm }^{l} = \sum_{\vec{l}}\sum_{i=1}^{3}( c_{\vec{l}+\vec{u}_i}^{\dagger}c_{\vec{l}} + {\rm H.c.} ) + \sum_{\vec{l}} U(\vec{l}) c_{\vec{l}}^{\dagger}c_{\vec{l}},
    \label{eqn:H_3DAA_infinite}
\end{equation}
where the points of the lattice are given by lattice vectors $\vec{l}=(l_1,l_2,l_3)$, or alternatively $\vec{l}=(l_x,l_y,l_z)$. The potential is given by
\begin{equation}
    U(\vec{l}) = 2W\sum_{i=1}^{3} \cos\Bigg( 2\pi \sum_{j=1}^{3} B_{ij}^{}l_{j}+\phi_i\Bigg).
\label{eqn:pot_infinite}
\end{equation}
The $B$-matrix gives the properties of the potential, which can become quasiperiodic w.r.t. the lattice. In the 1D Aubry-André model, it is known that the spectrum of the infinite size Hamiltonian is independent of the phase $\phi$, and thus it can be set to zero$~$\cite{avila2008}. We assume here that the same is true for the 3D Aubry-André model and set $\phi_i=0$ in the infinite system. If we choose the $B$-matrix to be symmetric, and we do so in our study, the Hamiltonian can be re-expressed using the dual Fourier transform
\begin{equation}
U_{\rm D}^f: c^{\dagger}_{\vec{k}^{\rm D}}=  L^{-3/2}\sum_{\vec{l}}e^{i 2\pi \vec{k}^{\rm D}\cdot B \cdot\vec{l}}c^{\dagger}_{\vec{l}},
    \label{eqn:H_3DAA_ftsd}
\end{equation}
where the product $\vec{k}^{\rm D}\cdot B \cdot\vec{l}$ is a number. In the dual space, the Hamiltonian has a form 
\begin{equation}
    H_{\rm }^{k^{\rm D}}= W\sum_{\vec{k}^{\rm D}}\sum_{i=1}^{3}( c_{\vec{k}^{\rm D}+\vec{u}_i}^{\dagger}c_{\vec{k}^{\rm D}}^{} + {\rm H.c.} ) + \sum_{\vec{k}^{\rm D}} U(\vec{k}^{\rm D}) c_{\vec{k}^{\rm D}}^{\dagger}c_{\vec{k}^{\rm D}}^{},
    \label{eqn:H_3DAA_infinite2}
\end{equation}
with the potential
\begin{equation}
    U(\vec{k}^{\rm D}) = 2\sum_{i=1}^{3} \cos\Bigg( 2\pi \sum_{j=1}^{3} B_{ij}^{}k^{\rm D}_{j}\Bigg).
\label{eqn:pot_infinite2}
\end{equation}
Here, the points of the dual momentum lattice are given by dual momentum lattice vectors $\vec{k}^{\rm D}=(k^{\rm D}_1,k^{\rm D}_2,k^{\rm D}_3)$, where $k^{\rm D}_i=1,\dots,L$. 

\subsection{$B$-matrix choice}
\label{choice bmatrix}

Here, we adopt the choice of $B$-matrix in Ref.$~$\cite{devakul2017anderson}, i.e., 
\begin{equation}
B=\beta
    \begin{bmatrix}   
    \mathrm{c}^2+\mathrm{s}^3 & \mathrm{c}\mathrm{s} & \mathrm{c}\mathrm{s}^2-\mathrm{c}\mathrm{s}\\
    \mathrm{c}\mathrm{s} & -\mathrm{s} & \mathrm{c}^2\\
    \mathrm{c}\mathrm{s}^2-\mathrm{c}\mathrm{s} & \mathrm{c}^2 & \mathrm{c}^2\mathrm{s}+\mathrm{s}^2
    \end{bmatrix},
\end{equation}
where $\mathrm{c}=\cos(\theta)$ and $\mathrm{s}=\sin(\theta)$ are chosen to depend on the same angle $\theta$, and the factor $\beta$ is typically chosen to be irrational. This matrix can be written as 
\begin{equation}
B=\beta
    \begin{bmatrix}
    -\mathrm{s} & 0 & \mathrm{c}\\
    0 & 1 & 0\\
    -\mathrm{c} & 0 & -\mathrm{s}
    \end{bmatrix}
    \begin{bmatrix}
    -\mathrm{s} & -\mathrm{c} & 0\\
    \mathrm{c} & -\mathrm{s} & 0\\
    0 & 0 & 1
    \end{bmatrix}
    \begin{bmatrix}
    -\mathrm{s} & 0 & \mathrm{c}\\
    0 & 1 & 0\\
    -\mathrm{c} & 0 & -\mathrm{s}
    \end{bmatrix}
    \begin{bmatrix}
    -1 & 0 & 0\\
    0 & 1 & 0\\
    0 & 0 & 1
    \end{bmatrix},
\end{equation}
which can be interpreted as a reflection followed by three Euler rotations by the same angle $\theta_1=\theta_2=\theta_3=\theta$. The geometrical interpretation of the transform by the matrix $B$ is that the incommensurate lattice defining the potential was originally aligned with the physical lattice. The former was subsequently reflected and rotated. 

We note that the particular choice of the $B$-matrix and the parameters $\beta$ and $\theta$ is one of all possible choices. In principle, for a simple $B$-matrix there exists a whole family of models given by the values of the pair $(\beta,\theta)$. Here, we adopt the choice of Ref.$~$\cite{devakul2017anderson}, i.e., $[\beta,\theta]=[(\sqrt{5}-1)/2,\pi/7]$.

\subsection{Commensurate approximation}
\label{commensurate approx}

We approach the behavior of the infinite system subjected to the incommensurate potential by considering a series of commensurate approximations with a commensurability of length $L$. The incommensurate quasiperiodic potential is replaced with an approximate potential that satisfies periodic boundary conditions on the cube of linear size $L$, i.e., with volume $V=L^3$, while retaining the dual properties of the infinite lattice. In practice, this is achieved by the replacement of $B$ with a suitable $B_{}^{\rm CA}$. Then, the eigenstates of the Hamiltonian in Eq.$~$\eqref{eqn:H_3DAA_infinite} with the potential from Eq.$~$\eqref{eqn:pot_infinite}, where $B$ is replaced by some $B_{}^{\rm CA}$, can be written in the form of a Bloch function,
\begin{equation}
    \Psi_{\vec{l}}^{}\propto e^{i\vec{h}\vec{l}}\psi_{\vec{l}}^{},
    \label{eqn:bloch_ansatze}
\end{equation}
where $h_i$ plays the role of the Bloch quasi-momentum from the Brillouin zone $h_i \in (-\pi/L,\pi/L]$.

The modulation parts of the Bloch function $\psi_{\vec{l}}^{}$ are given by the commensurate Hamiltonian defined in the cube of size $L$,
\begin{equation}
    H_{\rm CA}^{l,\phi,h} = \sum_{\vec{l}}\sum_{i=1}^{3}( e^{ih_i}c_{\vec{l}+\vec{u}_i}^{\dagger}c_{\vec{l}} + {\rm H.c.} ) + \sum_{\vec{l}} U_{{\rm CA}}^{\vec{\phi}}(\vec{l}) c_{\vec{l}}^{\dagger}c_{\vec{l}},
    \label{eqn:H_3DAA_real_comm}
\end{equation}
where the points of the lattice are given by lattice vectors $\vec{l}=(l_1,l_2,l_3)$, or alternatively $\vec{l}=(l_x,l_y,l_z)$, and where $l_i=1,\dots,L$. The commensurate potential is given by
\begin{equation}
    U_{{\rm CA}}^{\vec{\phi}}(\vec{l}) = 2W\sum_{i=1}^{3} \cos\Bigg( 2\pi \sum_{j=1}^{3} B_{ij}^{\rm CA}l_{j}+\phi_i\Bigg).
\label{eqn:pot_comm}
\end{equation}
We require that the Hamiltonian is periodic only at the size $L$. From this requirement, it follows that ${\rm det}(LB^{\rm CA})$ must be invertible mod $L$ and therefore coprime to $L$, i.e., $B^{\rm CA}$ fulfills the condition ${\rm gcd}({\rm det}(LB^{\rm CA}), L) = 1$, where ${\rm gcd}(n,m)$ is the greatest common divisor of $n$ and $m$. 

As we consider the commensurate approximations, the solutions $\psi_{\vec{l}}^{}$ depend on $h_i$ and $\phi_i$. The latter is restricted to the same Brillouin zone as $h_i$, i.e., $\phi_i \in (-\pi/L,\pi/L]$. Then, by studying the system-size scaling of properties of the Hamiltonian in Eq.$~$\eqref{eqn:H_3DAA_real_comm}, averaged over $\phi_i$ and $h_i$, we determine the properties of the incommensurate limit,
\begin{equation}
    \psi_{\vec{l}}^{}\xrightarrow{L\rightarrow \infty} \Psi_{\vec{l}}^{},
    \label{eqn:bloch_ansatze}
\end{equation}
i.e., since $h_i \xrightarrow{L\rightarrow \infty} 0$ and $\phi_i \xrightarrow{L\rightarrow \infty} 0$, $\psi_{\vec{l}}^{}$ become eigenstates of the Hamiltonian in Eq.$~$\eqref{eqn:H_3DAA_infinite}.

\subsection{Diagonal commensurate approximation}
\label{diag commensurate approx}

The approximation introduced in Ref.$~$\cite{devakul2017anderson}, dubbed the exactly commensurate approximation therein, can be in our framework understood as the commensurate approximation with the diagonal averaging $h_i=\phi_i$, i.e.,
\begin{equation}
    H_{\rm CA}^{l,\phi} \equiv H_{\rm CA}^{l,\phi,\phi} = \sum_{\vec{l}}\sum_{i=1}^{3}( e^{i\phi_i}c_{\vec{l}+\vec{u}_i}^{\dagger}c_{\vec{l}} + {\rm H.c.} ) + \sum_{\vec{l}} U_{\rm D}^{\vec{\phi}}(\vec{l}) c_{\vec{l}}^{\dagger}c_{\vec{l}},
    \label{eqn:H_3DAA_real_2}
\end{equation}
where the potential remains unchanged w.r.t. Eq.$~$\eqref{eqn:pot_comm}, 
\begin{equation}
    U_{\rm D}^{\vec{\phi}}(\vec{l}) = 2W\sum_{i=1}^{3} \cos\Bigg( 2\pi \sum_{j=1}^{3} B_{ij}^{\rm CA}l_{j} + \phi_i \Bigg).
\label{eqn:pot_self_mom_2}
\end{equation}
Since the full averaging of the Hamiltonian in Eq.$~$\eqref{eqn:H_3DAA_real_comm} and the diagonal averaging of the Hamiltonian in Eq.$~$\eqref{eqn:H_3DAA_real_2} give similar results close to the transition, see Appendix$~$\ref{fulldiag}, we consider only the later.

\subsection{$B$-matrix approximation choices}
\label{choice dual approx}
 
With the chosen $B$-matrix, we now turn to the construction of its commensurate approximation. This is not a unique construction. Here we explore two different approximations and show below that, apart from different fluctuations, they give the same results.

(i) A closest integer approximation. - The elements of $B^{\rm CA}$ are defined as
\begin{equation}
    (B^{\rm CA})_{ij}={\rm rnd}(LB^{}_{ij})/L,
\label{eqn:pot_sd_cl_int}
\end{equation}
where ${\rm rnd}()$ is the rounding of the number. Here, not all system sizes satisfy the co-prime condition. Thus, we consider only system sizes that do so.

(ii) A closest determinant approximation. -  We first construct the following sets of elements,
\begin{equation}
\{(B^{\rm CA})_{ij}\}=[{\rm floor}(LB_{ij})+\{0,1\}]/L,
\label{eqn:pot_sd_cl_det}
\end{equation}
i.e., for each element $(B^{\rm CA})_{ij}$ we consider the two closest integers ${\rm floor}(LB_{ij})$ and ${\rm floor}(LB_{ij})+1$.
In total, there are $2^{6}$ possible sets $\{(B^{\rm CA})_{ij}\}$, as there are six independent entries ${\rm floor}(LB_{ij})$ to which we add 0 or 1. Among these sets, we choose the set $\{(B^{\rm CA})_{ij}\}$ that minimizes a distance defined as
\begin{equation}
d=|1-|\det(B_{}^{\rm CA}/\beta)||,
\label{eqn:pot_sd_cl_det2}
\end{equation}
and which at the same time satisfies the co-prime condition. 

The distance $d$ measures how the absolute value of the determinant of $R_{}^{\rm CA}=B_{}^{\rm CA}/\beta$ deviates from 1. We note that for the original matrix $R=B/\beta$, the distance $d=|1-|\det(B_{}^{}/\beta)||=0$, as $R$ represents reflection followed by the Euler rotations, and thus $|\det(B_{}^{}/\beta)|=1$. We require the approximation to minimize this number such that it is closest to the original $B$-matrix. For all system sizes considered in our numerics, we can always find $B_{}^{\rm CA}$ satisfying the co-prime condition.

As can be seen from their construction, the commensurate approximations are slightly different for every choice of $L$. This may induce additional deviations on top of finite size effects, posing a challenge to the analysis of finite size scaling behavior. Despite these difficulties, in this section, we perform a finite-size scaling analysis and estimate the transition point. For both commensurate approximations presented above, we indeed observe some fluctuations in our measures when scaled with the system size $L$. However, after careful mitigation of these fluctuations, see Appendix$~$\ref{3D case fluctspec}, the main conclusions about the transition point in the thermodynamic limit are very similar, which suggests that the analysis is robust against these fluctuations. That being said, before the discussion of finite size scaling, we first discuss the dual symmetries of the model.

\subsection{Dual transform in finite cube}
\label{3D case transform}

The first step of the dual transform of the Hamiltonian in Eq.$~$\eqref{eqn:H_3DAA_real_2} consists of applying the unitary dual Fourier transform $U_{\rm D}^f$ to the Hamiltonian in Eq.$~$\eqref{eqn:H_3DAA_real_2}, with
\begin{equation}
U_{\rm D}^f: c^{\dagger}_{\vec{k}^{\rm D}}=L^{-3/2}\sum_{\vec{l}}e^{i 2\pi \vec{k}^{\rm D}\cdot B^{\rm CA} \cdot\vec{l}}c^{\dagger}_{\vec{l}},
    \label{eqn:H_3DAA_ftsd}
\end{equation}
where the product $\vec{k}^{\rm D}\cdot B^{\rm CA} \cdot\vec{l}$ is a number. Here, the points of the dual momentum lattice are given by dual momentum lattice vectors $\vec{k}^{\rm D}=(k^{\rm D}_1,k^{\rm D}_2,k^{\rm D}_3)$, where $k^{\rm D}_i=1,\dots,L$. 

We distinguish the dual Fourier transform from the standard Fourier transform defined as
\begin{equation}
U_{}^f : c_{k}^{\dagger}=L^{-3/2}\sum_{\vec{l}}e^{i 2\pi \vec{k}\cdot \frac{\mathbb{I}}{L} \cdot\vec{l} } c_{l}^{\dagger},
    \label{eqn:H_3DAA_ft}
\end{equation} where $\mathbb{I}$ is the $3 \times 3$ unit matrix. Here, the points of the standard reciprocal lattice are given by reciprocal lattice vectors $\vec{k}=(k_1,k_2,k_3)$, where $k_i=1,\dots,L$. 

The result of the unitary dual Fourier transform $U_{\rm D}^f$ transform applied to the commensurate approximation in Eq.$~$\eqref{eqn:H_3DAA_real_2} is
\begin{align}
\begin{split}
    H_{\rm CA}^{k,-\phi} =W\sum_{\vec{k}}\sum_{i=1}^{3}( e^{-i\phi_i}c_{\vec{k}^{\rm D}+\vec{u}_i}^{\dagger}c_{\vec{k}^{\rm D}} + {\rm H.c.} ) \\
    + 2\sum_{i=1}^{3} \cos\Bigg( 2\pi \sum_{j=1}^{3} B_{ij}^{\rm CA}k^{\rm D}_{j} + \phi_i \Bigg)c_{\vec{k}}^{\dagger}c_{\vec{k}}.
    \label{eqn:H_3DAA_mom_0}
\end{split}
\end{align}
The second step of the dual transform is reversing the flux $\phi_i \rightarrow -\phi_i$ in Eq.$~$\eqref{eqn:H_3DAA_mom_0} for each direction $i$, while keeping the phase shift in the potential in Eq.$~$\eqref{eqn:pot_self_mom_2} the same. The result of the dual transform is
\begin{align}
\begin{split}
    H_{\rm CA}^{k,\phi} =W\sum_{\vec{k}}\sum_{i=1}^{3}( e^{i\phi_i}c_{\vec{k}^{\rm D}+\vec{u}_i}^{\dagger}c_{\vec{k}^{\rm D}} + {\rm H.c.} ) \\
    + 2\sum_{i=1}^{3} \cos\Bigg( 2\pi \sum_{j=1}^{3} B_{ij}^{\rm CA}k^{\rm D}_{j} + \phi_i \Bigg)c_{\vec{k}}^{\dagger}c_{\vec{k}}.
    \label{eqn:H_3DAA_mom}
\end{split}
\end{align}
We can summarize the dual transform with
\begin{equation}
    H_{\rm CA}^{l,\phi}[1,2W]  \xrightarrow{(U_{\rm D}^f)^\dagger H U_{\rm D}^f} H_{\rm CA}^{k,-\phi}[W,2] \rightarrow   H_{\rm CA}^{k,\phi}[W,2],
     \label{eqn:dual_trans}
\end{equation}
where the first and second terms in brackets $[.,.]$ refer to the coefficients in front of the kinetic and potential terms, respectively.

\subsection{Unitary transform in finite cube and boundary twist}
\label{3D case transform twist}

One can ask whether the second step of Eq.$~$\eqref{eqn:dual_trans} could be expressed in terms of a unitary transform. To this end we locally gauge the creation operators,
\begin{align}
\begin{split}\label{eqn:H_3DAA_phase_twist}
U_{\phi}: e^{i2\vec{k}^{\rm D}\vec{\phi}}\tilde{c}^{\dagger}_{\vec{k}^{\rm D}}= c^{\dagger}_{\vec{k}^{\rm D}},
\end{split}
\end{align}
and similarly for the annihilation operator.
This effectively switches the Peierls phases in the bulk of the cube,
\begin{align}
\begin{split}
&e^{-i\phi_i}c^{\dagger}_{\vec{k}^{\rm D}+\vec{u}_i}c^{}_{\vec{k}^{\rm D}}=e^{i\phi_i}\tilde{c}^{\dagger}_{\vec{k}^{\rm D}+\vec{u}_i}\tilde{c}^{}_{\vec{k}^{\rm D}}
    \label{eqn:H_3DAA_ftsd_twist0}
\end{split}
\end{align}
but it creates a twist at the boundary
\begin{align}
\begin{split}
&e^{-i\phi_i}c^{\dagger}_{(..,L-1,..)+\vec{u}_i}c^{}_{(..,0,..)}=\\
&e^{i\phi_i}e^{i2\phi_i(L-1)}\tilde{c}^{\dagger}_{(..,L-1,..)+\vec{u}_i}\tilde{c}^{}_{(..,0,..)}.
    \label{eqn:H_3DAA_twist1}
\end{split}
\end{align}
The whole transformation can be then summarized as
\begin{equation}
     H_{\rm CA}^{l,\phi}[1,2W]  \xrightarrow{(U_{\rm D}^f)^\dagger H U_{\rm D}^f} H_{\rm CA}^{k,-\phi}[1,2W] \xrightarrow{(U_{\phi})^\dagger H U_{\phi}}  \tilde{H}_{\rm CA}^{k,\phi}[W,2],
     \label{eqn:dual_trans_twist}
\end{equation}
where, $\tilde{H}_{\rm CA}^{k,\phi}[W,2]$ has the same form  in the bulk as the Hamiltonian $H_{\rm CA}^{k,\phi}[W,2]$ in Eq.$~$\eqref{eqn:dual_trans}, but compared to that, it has a twisted boundary.
We will use this transform to argue about the dual symmetry of particle transport in Sec.$~$\ref{3D case time overlap}.

\subsection{Symmetry of the eigenspectra}\label{3D case duality spectra}

\begin{figure}[t]
\includegraphics[width=0.45\textwidth]{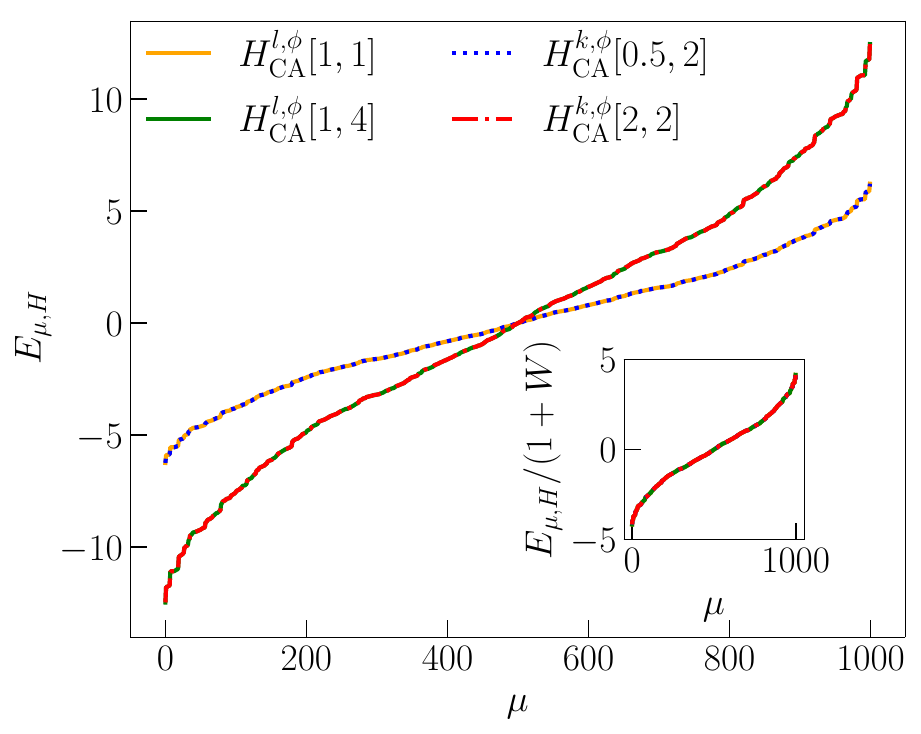}
  \caption{
  Example of spectra of the 3D Aubry-André model, for $L=10$, $\phi_x, \phi_y, \phi_z=0.1/L$, $\theta=\pi/7$ and $\beta=(\sqrt{5}-1)/2$ in the closest integer approximation of the $B$-matrix.
  Inset: The rescaled energy spectrum from the main panel by factor $1/(1+W)$.
  }
  \label{fig:3DAA_dual_example}
\end{figure}

We now consider the eigenspectrum $E_{\mu,H}$ of the Hamiltonian, which is defined as
\begin{equation}\label{eqn:schrodinger}
   H|\mu_H\rangle=E_{\mu,H}|\mu_H\rangle.
\end{equation}
The dual transform, as it consists of a unitary transform and complex conjugation, does not change the spectrum of the Hamiltonian. Thus, the Hamiltonians $H_{\rm CA}^{l,\phi}[1,2W]$ and $H_{\rm CA}^{k,\phi}[W,2]$ share the same eigenspectrum and we can write 
\begin{equation}
    E_{\mu,H} \{H_{\rm CA}^{l,\phi}[1,2W]\}= E_{\mu,H} \{H_{\rm CA}^{k,\phi}[W,2]\}.
     \label{eqn:spectrum1}
\end{equation}
As a check of our implementation, this trivial relation is demonstrated in the main panel of Fig.$~$\ref{fig:3DAA_dual_example}. In the first case, we directly diagonalize $H_{\rm CA}^{l,\phi}[1,2W]$. In the second case, we first implement the transform in Eq.$~$\eqref{eqn:dual_trans} and then diagonalize $H_{\rm CA}^{k,\phi}[W,2]$.

We can further divide the Hamiltonian in Eq.$~$\eqref{eqn:H_3DAA_mom} by $W$, which gives us
\begin{align}\nonumber
    &
    H_{\rm CA}^{k,\phi}[1,2/W] =
    \sum_{\vec{k}^{\rm D}}\sum_{i=1}^{3}( e^{i\phi_i}c_{\vec{k}^{\rm D}+\vec{u}_i}^{\dagger}c_{\vec{k}^{\rm D}} + {\rm H.c.} ) +   
    \\ &  2/W\sum_{i=1}^{3} \cos\Biggl( 2\pi \sum_{j=1}^{3} B_{ij}^{\rm CA}l_{j} + \phi_i \Biggl) = \frac{H_{\rm CA}^{k,\phi}[W,2]}{W}.
    \label{eqn:H_3DAA_real_block_dual2}
\end{align}
Similarly, we can write down the Hamiltonian in the lattice basis,
\begin{align}\nonumber
    &
    H_{\rm CA}^{l,\phi}[1,2/W] =  \sum_{\vec{k}^{\rm D}}\sum_{i=1}^{3}( e^{i\phi_i}c_{\vec{l}^{}+\vec{u}_i}^{\dagger}c_{\vec{l}^{}} + {\rm H.c.} ) +   
    \\ &  2/W\sum_{i=1}^{3} \cos\Biggl( 2\pi \sum_{j=1}^{3} B_{ij}^{\rm CA}l_{j} + \phi_i \Biggl).
    \label{eqn:H_3DAA_real_block_dual3}
\end{align}
Now, as the Hamiltonian in Eq.$~$\eqref{eqn:H_3DAA_real_block_dual2} has the same form as the one in Eq.$~$\eqref{eqn:H_3DAA_real_block_dual3}, they share the same spectrum, i.e.,
\begin{equation}
    E_{\mu,H} \{H_{\rm CA}^{k,\phi}[1,2/W]  \}= E_{\mu,H} \{H_{\rm CA}^{l,\phi}[1,2/W]\}.
     \label{eqn:spectrum2}
\end{equation}
Finally, we can put together Eq.$~$\eqref{eqn:spectrum1}, Eqs.$~$\eqref{eqn:H_3DAA_real_block_dual2}, and$~$\eqref{eqn:spectrum2} to derive the spectrum symmetry
\begin{align}\nonumber
    &
    \frac{E_{\mu,H} \{H_{\rm CA}^{l,\phi}[1,2W]\}}{W}= \frac{E_{\mu,H} \{H_{\rm CA}^{k,\phi}[W,2]\}}{W}\\
    &=E_{\mu,H} \{H_{\rm CA}^{k,\phi}[1,2/W]  \}= E_{\mu,H} \{H_{\rm CA}^{l,\phi}[1,2/W]\}.
     \label{eqn:spectrum_connection}
\end{align}

The first and last terms of Eq.$~$\eqref{eqn:spectrum_connection} give us
\begin{align}
    \frac{E_{\mu,H} \{H_{\rm CA}^{l,\phi}[1,2W]\}}{W}= E_{\mu,H} \{H_{\rm CA}^{l,\phi}[1,2/W]\},
     \label{eqn:spectrum_connection2}
\end{align}
which can be further re-written in symmetrical form as 
\begin{align}
    \frac{E_{\mu,H} \{H_{\rm CA}^{l,\phi}[1,2W]\}}{1+W}= \frac{E_{\mu,H} \{H_{\rm CA}^{l,\phi}[1,2/W]\}}{1+1/W}.
     \label{eqn:spectrum_connection3}
\end{align}
Equation$~$\eqref{eqn:spectrum_connection3} is the basic symmetry of the spectrum of the Hamiltonian in Eq.$~$\eqref{eqn:H_3DAA_real_2}. The spectra for parameters $(\phi,W)$ and $(\phi',W')=(\phi,1/W)$ are the same when divided by the factors $1+W$ and $1+W'=1+1/W$, respectively.  We illustrate this relation with an example of rescaled spectra for $W=0.5$ and $W'=1/W=2$ in the inset of Fig.$~$\ref{fig:3DAA_dual_example}. Indeed, we observe a collapse of the spectra when rescaled by the factors $1/1.5$ and $1/3$, respectively. 

\subsection{Symmetry of the eigenstate amplitudes and inverse participation ratio}  \label{3D case eigenstates}

\begin{figure}[t]
\includegraphics[width=0.45\textwidth]{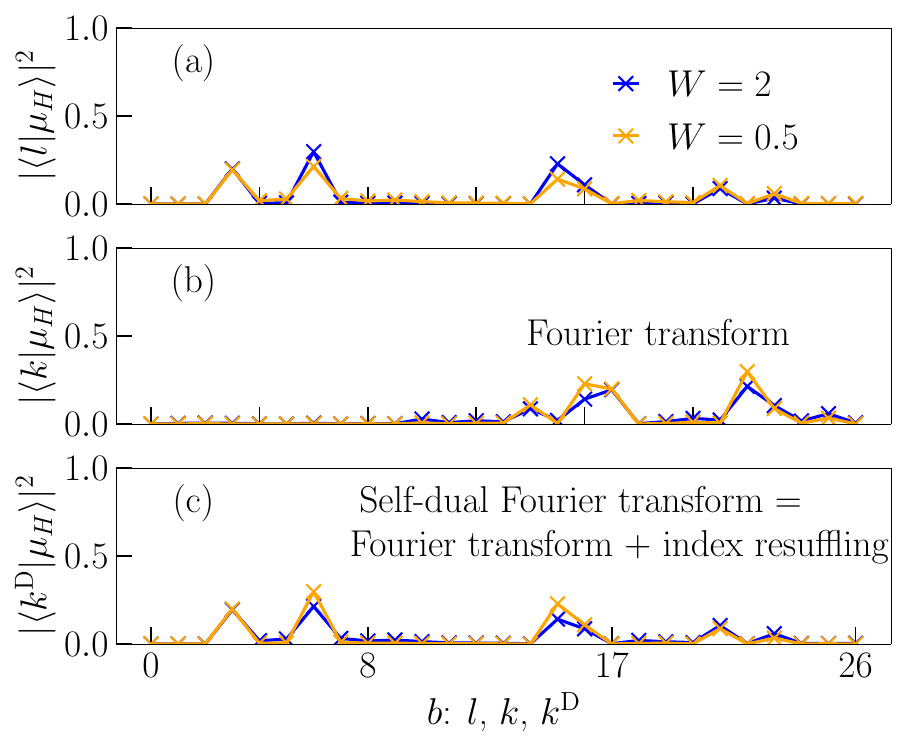}
  \caption{Examples of ground state amplitude profiles in 3D Aubry-André model, for $L=3$, $\phi_x, \phi_y, \phi_z=0.1/L$, $\theta=\pi/7$ and $\beta=(\sqrt{5}-1)/2$ in the closest integer approximation of the $B$-matrix.
  The amplitudes are considered in (a) the real-space basis $|b\rangle=|l\rangle$, (b) the standard momentum basis $|b\rangle=|k\rangle$, and (c) the dual momentum basis $|b\rangle=|k^{\rm D}\rangle$.
  }
  \label{fig:3DAA_dual_example2}
\end{figure}

We now study the properties of the eigenstates $|\mu\rangle$ of the Hamiltonian $H_{\rm CA}^{l,\phi}$ in Eq.$~$\eqref{eqn:H_3DAA_real_2}. We study the eigenstate amplitudes $|\langle b|\mu_H\rangle|^2$ in three different bases, where the considered bases $|b\rangle$ are
\begin{align}\nonumber
&|l\rangle=c_{\vec{l}}^{\dagger}|\varnothing\rangle,\\\nonumber
&|k\rangle=c_{\vec{k}}^{\dagger}|\varnothing\rangle=L^{-3/2}\sum_{\vec{l}} e^{i 2\pi \vec{l} \frac{1}{L} \vec{k}}c_{\vec{l}}^{\dagger}|\varnothing\rangle,\\ \label{eq:basis}
&|k^{\rm D}\rangle=c^{\dagger}_{\vec{k}^{\rm D}}|\varnothing\rangle=L^{-3/2}\sum_{\vec{l}}e^{i 2\pi \vec{k}^{\rm D}\cdot B^{\rm CA} \cdot\vec{l}}c^{\dagger}_{\vec{l}}|\varnothing\rangle.
\end{align}
We define the indices $l=l_x+l_yL+l_zL^{2}$ in the real-space basis, $k=k_x+k_yL+k_zL^{2}$ in the standard Fourier basis, and $k^{\rm D}=k^{\rm D}_x+k^{\rm D}_yL+k^{\rm D}_zL^{2}$ in the dual Fourier basis.

As an example, in Fig.$~$\ref{fig:3DAA_dual_example2}(a), we show the amplitudes of ground states for $W=0.5$ and $W'=1/W=2$ as a function of the index $l$. We observe that the ground state is delocalized for both $W=0.5$ and $W'=2$ with a slightly higher degree of localization for $W'=2$.

In the standard Fourier space, Fig.$~$\ref{fig:3DAA_dual_example2}(b), the situation is opposite, the ground state is more delocalized for $W'=2$ than for $W=0.5$. However, the profile of the eigenstates in the $k$ basis is different from the profile in the $l$ basis. To obtain the correct position in dual space, the dual Fourier transform should be used instead, see Eq.$~$\eqref{eqn:H_3DAA_ftsd} and the last line in Eq.$~$\eqref{eq:basis}. Indeed, in Fig.$~$\ref{fig:3DAA_dual_example2}(c) we observe that the profiles of the ground states in the dual momentum space for $W'=2$ and $W=0.5$ are exactly the same as the profiles of the ground states in the original lattice space for $W=0.5$ and $W'=2$, respectively; compare Fig.$~$\ref{fig:3DAA_dual_example2}(a) with Fig.$~$\ref{fig:3DAA_dual_example2}(c).

\begin{figure}[t]
\includegraphics[width=0.45\textwidth]{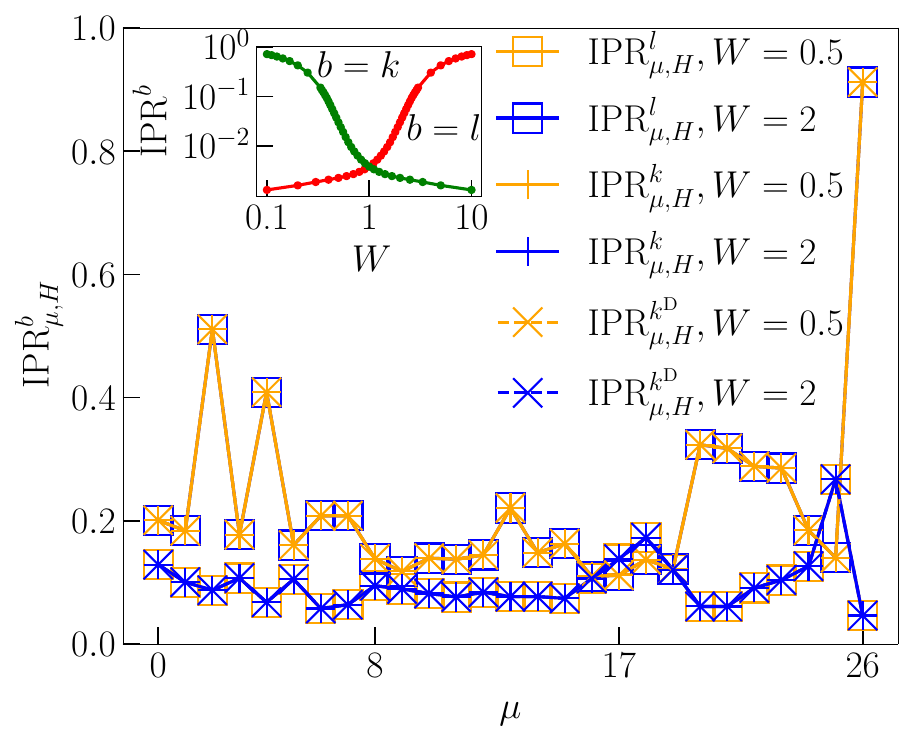}
  \caption{Examples of ${\rm IPR}_\mu^b$ in 3D Aubry-André model, for $L=3$, $\phi_x, \phi_y, \phi_z=0.1/L$, $\theta=\pi/7$ and $\beta=(\sqrt{5}-1)/2$ in the closest integer approximation of the $B$-matrix. The IPR is considered in the real-space basis $|b\rangle=\{ |l\rangle\}$, in the standard momentum basis $|b\rangle=\{|k\rangle\}$, and in the dual momentum basis $|b\rangle=\{|k^{\rm D}\rangle\}$,  respectively. Inset: Symmetry of averaged ${\rm IPR}^l$ (red) and ${\rm IPR}^k$ (green) as functions of $W$. We average over $50\%$ of the bulk eigenstates, i.e., $N_i=V/4$ and $N_C=V/2$, and over 100 realization of $\{\phi_x,\phi_y,\phi_z\}$.} 
  \label{fig:3DAA_dual_approximation_iprs}
\end{figure}

To quantify the degree of localization in a given basis, we use the inverse participation ratio (IPR). The IPR of an eigenstate $|\mu_{H}\rangle$ in a basis $|b\rangle$ is defined as
\begin{equation}
    {\rm IPR}_{\mu,H}^{b} = \sum_{b=1}^{L^{3}} |c_{\mu,H}^b|^{4}.
      \label{eq:IPR_def}
\end{equation}
where $|\mu_{H}\rangle=\sum_{b=1}^V c_{\mu,H}^b|b\rangle$, with $c_{\mu,H}^b=\langle b|\mu_H\rangle$ being the eigenstate coefficients. Here, we consider the relevant bases $|b\rangle=\{|l\rangle, |k\rangle, |k^{\rm D}\rangle\}$, see Eq.$~$\eqref{eq:basis}. In Fig.$~$\ref{fig:3DAA_dual_approximation_iprs}, we show examples of ${\rm IPR}_{\mu,H}^l$, ${\rm IPR}_{\mu,H}^k$, and ${\rm IPR}_{\mu,H}^{k^{\rm D}}$ for all eigenstates of the Hamiltonian for which the ground state was discussed in Fig.$~$\ref{fig:3DAA_dual_example2}. 

First, the results demonstrate that the IPR has the same value in the standard and in the dual momentum basis, ${\rm IPR}_{\mu,H}^{k_{}}={\rm IPR}_{\mu,H}^{k^{\rm D}}$, i.e., the IPR is independent of how the indices are shuffled between $|k\rangle$ and $|k^{\rm D}\rangle$. Additionally, the IPR in the lattice basis for $W$ is the same as the IPR in the momentum basis for $1/W$, i.e., ${\rm IPR}_{\mu,H_{\rm CA}^{l,\phi}[1,2W]}^{l} ={\rm IPR}_{\mu,H_{\rm CA}^{l,\phi}[1,2/W]}^{k^{\rm D}}$. This is the basic symmetry of the eigenstates.

\begin{figure}[t]
    \includegraphics[width=0.45\textwidth]{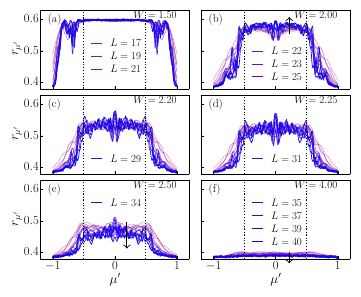}
  \caption{The eigenstate-resolved gap ratio $r_{\mu'}$ as a function of the scaled indices $\mu'= (\mu-\frac{V}{2})/\frac{V}{2}$ for the 3D Aubry-André model for $\theta=\pi/7$ and $\beta=(\sqrt{5}-1)/2$ in the closest integer approximation of the $B$-matrix. The two vertical dash-dotted lines correspond to the $50\%$ of energy levels around the middle of the spectrum. The arrows in (b), (e) and (f) indicate the flow of the plateaus with the system size. We average over 100 Hamiltonian realizations.}
  \label{fig:r_ratio}
\end{figure}

Finally, we study the average IPR defined as
\begin{equation}
    {\rm IPR}^b
    = \Bigg\langle \frac{1}{N_{C}} \sum_{\mu=N_{{\rm i}}}^{N_{{\rm i}} + N_{C}-1} {\rm IPR}_{\mu,H}^{b}\Bigg\rangle_H.
    \label{eq:IPR average} 
\end{equation} 
where we average over an energy band of $N_{C}$ states, and $\langle...\rangle_H$ stands for the average over $N_R$ Hamiltonian realizations (realized by different phases).

In the inset of Fig.$~$\ref{fig:3DAA_dual_approximation_iprs}, we show
examples of ${\rm IPR}^l$ and ${\rm IPR}^{k^{\rm D}}$
as functions of $W$. As expected, there is reflection symmetry of the IPR 
\begin{equation}
{\rm IPR}^l[W] = {\rm IPR}^k[1/W].
\label{eq:IPR average_symmetry} 
\end{equation}
Considering this symmetry, in the following sections, it is sufficient to focus only on the IPR of the original real-space lattice, i.e., ${\rm IPR}^l$, as the properties of ${\rm IPR}^k$ can be easily deduced from the symmetry.

\subsection{Edge effects and level spacing ratio}  \label{3D case edges}

It is well known$~$\cite{Schubert05} that in the paradigmatic 3D disordered Anderson model the transition does not occur for all eigenstates at the same disorder strength. The Anderson model exhibits mobility edges, i.e., delocalized eigenstates in the middle of the spectrum  separated from the localized eigenstates at the edge of the spectrum. We therefore ask whether this scenario is also relevant for the $3$D Aubry-André model.

To this end, we consider the energy dependence of the spectral properties, and we study the statistics of the gap ratio$~$\cite{oganesyan_huse_07}, defined as
\begin{equation}\label{eq:gap_ratio_def}
    r_{\mu,H} = \frac{{\rm min}(s_{\mu,H}, s_{\mu+1,H})}{{\rm max}(s_{\mu,H}, s_{\mu+1,H})}.
\end{equation}
where 
\begin{equation}\label{eq:gap_def}
s_{\mu,H} = E_{\mu+1,H}-E_{\mu,H}
\end{equation}
are the gaps between the two consecutive eigenenergy levels $E_{\mu,H}$ and $E_{\mu+1,H}$ of the Hamiltonian $H$, see Eq.$~$\eqref{eqn:schrodinger}.

To address the question of mobility edges, we study the eigenstate-resolved gap ratio,
\begin{equation}\label{eq:gap ratio average_mu} 
    r_{\mu}= \langle r_{\mu,H}\rangle_H,
\end{equation} 
averaged over the Hamiltonian realizations. In order to compare different system sizes we plot $r_{\mu}$ as a function of rescaled indices $\mu'= (\mu-\frac{V}{2})/\frac{V}{2}$, such that the indices lie in the range $[-1,1]$. Thus, $\mu'=-1$ represents the bottom of the band and $\mu'=1$ is the top of the band. We note that the single-particle bandwidth does not increase with the system size, as the $3$D Aubry-André model is quadratic.

In Fig.$~$\ref{fig:r_ratio}, we plot $r_{\mu'}$ for a few representative values of $W$ and several system sizes $L$. We observe that the gap ratio is constant, i.e., it shows a plateau with finite-size fluctuations for roughly $50\%$ of energy levels around the middle 
of the spectrum. We denote the approximate range of the plateau with dashed vertical lines in Fig.$~$\ref{fig:r_ratio}. The states outside of this mid-spectral range show a higher degree of localization than the states in the mid-spectral range. 

On the one hand, for some values of the potential strength, see Figs.$~$\ref{fig:r_ratio}(a) and$~$\ref{fig:r_ratio}(b), the plateau value tends to flow with the system size towards the value given by the GUE. On the other hand, for some values of the potential strength, it instead decreases to the Poisson value, see Figs.$~$\ref{fig:r_ratio}(e) and$~$\ref{fig:r_ratio}(f). For values in the range $W=2.18-2.25$, the plateau appears rather stable, see Figs.$~$\ref{fig:r_ratio}(c) and$~$\ref{fig:r_ratio}(d). This indicates a possible transition point between the GUE and the Poisson behavior. 

Outside of the plateau region, i.e., $\mu'\rightarrow -1$ and $\mu'\rightarrow 1$, the values of $r_{\mu'}$ do not seem to increase to the GUE value. Rather, they decrease. We interpret them as edge states, which are probably localized in the thermodynamic limit. This indicates that the $3$D Aubry-André model indeed exhibits localized edges, similarly to the paradigmatic 3D disordered Anderson model. Another aspect of the 3D Aubry-André model that can be inferred from Fig.$~$\ref{fig:r_ratio} are rather strong fluctuations of the value of $r_{\mu}$ around the plateaus. The fluctuations are induced by imposing the dual conditions of the model. This is very different from the 3D Anderson model, where such strong fluctuations are not present$~$\cite{jiricek2026informationcompressioncriticality}. In the next sections, we discuss how to cope with the fluctuations so that the transition can still be characterized with high precision.

\section{Detection of the eigenstate transitions} \label{3D case}

\subsection{Finite-size scaling analysis of $r$ and IPR}
 \label{3D case scaling}

Our goal is to determine the critical value $W_c^{}$ from a finite-size scaling analysis of the average gap ratio and IPR. However, the strong fluctuations observed in Fig.$~$\ref{fig:r_ratio} pose a challenge to standard procedures for characterizing eigenstate transitions. For example, in studies of the transition in the 3D Anderson model, a vanishing fraction of eigenstates is typically considered, see, e.g., Ref.$~$\cite{suntajs_prosen_21}. However, such averaging cannot be applied to the 3D Aubry-André model. If we consider the average over the vanishing fraction, there would be strong fluctuations of the averaged value with system size, showing practically no crossing point in $r$ as a function of $W$. Consequently, this would hinder any estimate of the transition point.  

Here, we need to resort to rather unconventional averaging over the non-vanishing portion of eigenstates. Based on the analysis of results in Fig.$~$\ref{fig:r_ratio}, we found that averaging over the $50\%$ states is a good strategy for the present model. It can significantly reduce the fluctuations of the averaged quantity, however, the effects of the mobility edges are excluded. Therefore, in the rest of  Sec.$~$\ref{3D case}, we focus on $50\%$ of the eigenstates in the middle of the spectra. This allows for precise determination the eigenstate transition.

Another unconventional feature of the finite size-scaling in the present model is the choice of ansatz for the correlation length function. Here, due to the symmetry of the results, the ansatz should describe the vicinities of $W_c$ and $1/W_c$ on equal footing. In this respect, the standard ansatz, i.e., $\xi \approx (1/|W-W_c|)^{\nu}$ with the exponent $\nu$, fails to do so. To resolve this issue, we apply the ansatz for the correlation length inspired by the original work of Aubry and André$~$\cite{Aubry80}. They argued that the correlation/localization length in 1D is given by $\xi=\big(\frac{1}{|\ln(W/W_c)|}\big)^{\nu}$, with $W_c^{\rm 1D}=1$ and $\nu^{\rm 1D}=1$. 

We adopt the ansatz of Aubry and André for the 3D Aubry-André model; however, we consider the transition point $W_c^{}$ and the exponent $\nu$ as fitting parameters to be determined. In this way, the symmetry of the model is retained. We note that the symmetric ansatz can be related to the standard formulation as follows. The logarithm, rewritten as $\ln[1+(W-W_c)/W_c]$, can be expanded to the first order, giving $1/\ln[1+(W-W_c)/W_c] \approx W_c/|W-W_c|$. Thus, to the first order, $\xi \approx 1/|W-W_c|$. 

\begin{figure}[t]
  \includegraphics[width=0.5\textwidth]{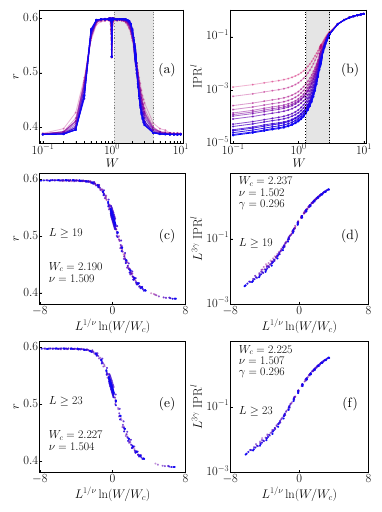}
  \caption{(a) Average gap ratio $r$ and (b) average IPR, versus $W$, for the 3D Aubry-André model for $\theta=\pi/7$ and $\beta=(\sqrt{5}-1)/2$ in the closest integer approximation of the $B$-matrix. We average over $50\%$ of the eigenstates around the mid-spectra and over $100$ Hamiltonian realizations. The color code denoting the systems sizes is the same as in Fig.$~$\ref{fig:r_ratio}.
    (c) and (e) Collapse of $r$ and, (d) and (f) collapse of IPR, following from the cost function minimization. We use the values in the shaded region of panels (a) and (b) to perform the collapse for $r$ and IPR, respectively.}
  \label{fig:r_ipr_collapse2}
\end{figure}

Having discussed the unconventional features of finite size scaling in the present model, we are ready to estimate the fitting parameters, $W_c^{}$ and $\nu$. To do so, we use the cost function minimization procedure of Ref.$~$\cite{suntajs_bonca_20a}. The cost function for a quantity $X$, which consists of $N_{{\rm p}}$ values, can be written as
\begin{equation}
    C_{{\rm X}} = \frac{ \sum_{j=1}^{N_{{\rm p}}-1} | X_{j+1} - X_{j} | }{ {\rm max}\{ X_j \} - {\rm min}\{ X_j \} } - 1.
    \label{eqn:cost}
\end{equation}
Each value $X_j$ is the quantity of interest, e.g. the averaged gap ratio, for various values $W$ and $L$. In Eq.$~$\eqref{eqn:cost}, we sort all $N_{{\rm p}}$ values of $X_j$ according to nondecreasing values of $(L/\xi)^{1/\nu}=L^{1/\nu}\ln(W/W_{{\rm c}})$. In the case of an ideal data collapse, this implies $\sum_{j}|X_{j+1} - X_{j}| = {\rm max}\{ X_j \} - {\rm min}\{ X_j \}$ and hence $C_X=0$. For real data, the cost function is always positive, $C_X>0$. Our goal is to find solutions that minimize $C_X$. In order to obtain the best value of the transition point, a large data set is needed. We could, in principle, include all the datasets for all system sizes studied here.  However, since they are likely to be well outside the scaling regime, the inclusion of data for small system sizes lowers the quality of the fit of the parameters. Thus, a compromise between including as many data sets as possible and excluding the finite size effects needs to be made. For example, for the system sizes of the closest integer approximation considered in this study, i.e., $L\in\{10,11,13,17,19,21,22,23,25,29,31,34,35,37,39,40\}$,
we found that the collapses of our results considering system sizes $L\geq 19$, $L\geq 21$, $L\geq 22$, and $L\geq 23$, are reasonable compromises, giving internally consistent estimates of the transition. In Fig.$~$\ref{fig:r_ipr_collapse2}, we show examples of such collapses of our results considering system sizes $L\geq 19$ and $L\geq 23$.

We first study the average gap ratio $r$, defined as
\begin{equation}
    r= \Bigg\langle \frac{1}{N_{C}-2}\sum_{\mu=N_{{\rm i}}}^{N_{{\rm i}} + N_{C}-3} r_{\mu,H}\Bigg\rangle_H,
        \label{eq:gap ratio average} 
\end{equation} 
where, analogously to the IPR in Eq.$~$\eqref{eq:IPR average}, we average over an energy band of $N_{C}$ states, and $\langle...\rangle_H$ means an average over $N_R$ Hamiltonian phase realizations. Based on the discussion in the beginning of this section, we set $N_i=V/4$ and $N_C=V/2$ in Eqs.$~$\eqref{eq:gap ratio average} and$~$\eqref{eq:IPR average}. 

In Fig.$~$\ref{fig:r_ipr_collapse2}(a), we show $r$ as a function of $W$ for several system sizes. As a side remark, we note that the same results as in Fig.$~$\ref{fig:r_ipr_collapse2}(a) are also used in Fig.$~$\ref{fig:intro} in the introduction. We observe that the $50\%$-averaging provides smooth curves that exhibit a crossing at values in the range $W=2.18-2.25$, and the curves serve as a good basis for the finite-size scaling analysis. 

We therefore focus on the vicinity of $W\approx 2.2$, as seen in the gray area in  Fig.$~$\ref{fig:r_ipr_collapse2}(a), which indicates the range of values of $W\in[1.1,4.0]$ considered for the finite-size analysis. Examples of results of the cost function minimization are shown in Figs.$~$\ref{fig:r_ipr_collapse2}(c) and$~$\ref{fig:r_ipr_collapse2}(e) for data sets that include all system sizes $L\geq 19$ and $L\geq 23$, respectively. We observe good collapses with the fit of the transition point $W_c^{} \approx 2.19-2.23$, and the fit of the correlation length exponent $\nu \approx 1.504-1.509$.

The results for the average IPR are shown in Fig.$~$\ref{fig:r_ipr_collapse2}(b). The IPR exhibits a decay with system size $L$ for values $W\lessapprox2.3$ and a saturation with system size $L$ for higher values of $W$. The IPR decay is assumed to follow a power-law ${\rm IPR} \propto V^{-\gamma}=L^{-3\gamma}$, and at the transition point, the decay is expected to be governed by the fractal dimension $\gamma<1$ which is unknown. Thus, in the case of IPR, the quantity to be collapsed is $X=V^{\gamma}{\rm IPR}=L^{3\gamma}{\rm IPR}$. The fractal dimension $\gamma$ is an additional fitting parameter to be determined. The gray area in Fig.$~$\ref{fig:r_ipr_collapse2}(b) indicates the range of values of $W\in[1.3,3.0]$ considered for the finite-size analysis of the IPR.

Examples of results of the cost function minimization  are shown in Figs.$~$\ref{fig:r_ipr_collapse2}(d) and$~$\ref{fig:r_ipr_collapse2}(f) for data sets that include all system sizes $L\geq 19$ and $L\geq 23$, respectively. We observe quite good collapses with the fit of the transition point $W_c^{} \approx 2.22-2.23$ and the fit of the correlation length exponent $\nu \approx 1.502-1.507$. 

We note that despite the different implementations in Refs.$~$\cite{devakul2017anderson,Sutradhar19,Luo22}, the estimates of the critical point $W_c$ and the correlation length exponent $\nu$ therein, i.e., $W_c\approx 2.17$ and $\nu \approx 1.61$$~$\cite{devakul2017anderson}, $W_c\approx 2.22 \pm0.01$ and $\nu \approx 1.6 \pm0.04$$~$\cite{Sutradhar19}, and $W_c\approx 2.229-2.233$ and $\nu \approx 1.599-1.613$$~$\cite{Luo22}, are consistent with our estimate. 

\begin{figure}[t]
  \includegraphics[width=0.45\textwidth]{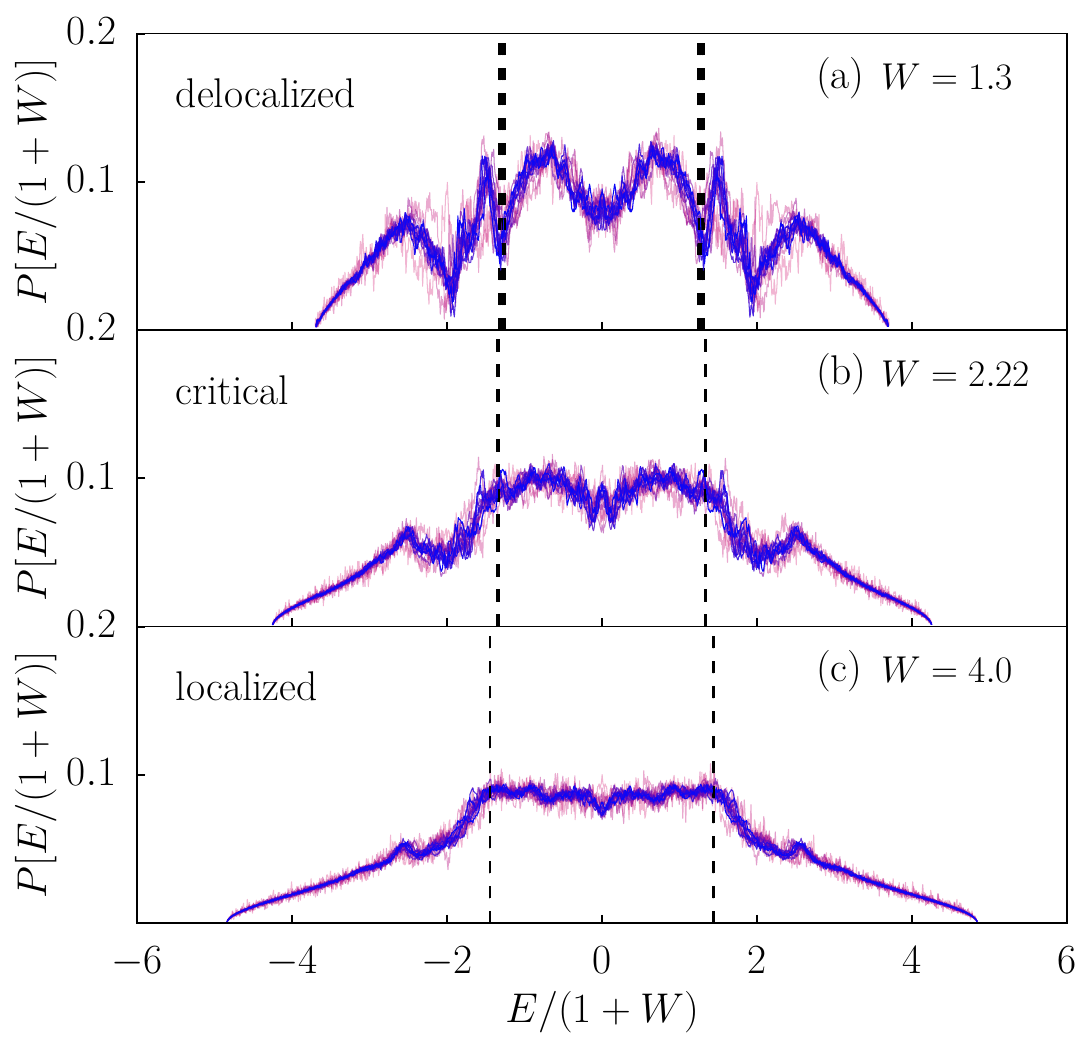}
  \caption{Density of states for the 3D Aubry-André model for $\theta=\pi/7$ and $\beta=(\sqrt{5}-1)/2$ in the closest integer approximation of the $B$-matrix. We average over $50\%$ of the eigenstates around the mid-spectra and over $100$ Hamiltonian realizations. The color code denoting the systems sizes is the same as in Fig.$~$\ref{fig:r_ratio}. The vertical dash-dotted lines correspond to the $50\%$ of energy levels around the middle of the spectrum.}
  \label{fig:dos_3daa}
\end{figure}

Finally, our scaling collapses suggest that the value of the transition point is within the range $W_c^{}\approx 2.19-2.23$. However, the finite size scaling analysis cannot provide a more accurate determination, as the curves are still affected by systematic errors induced by the imposition of the commensurate approximation $B^{\rm CA}$ in Eq.$~$\eqref{eqn:pot_self_mom_2}. We discuss the nature of the systematic errors and provide an improved determination of the transition point in Sec.$~$\ref{3D case universal}. Before doing so, in Sec.$~$\ref{3D case spectral form factor} and Sec.$~$\ref{3D case filter}, we study in detail the properties of the system deep in the ergodic phase ($W$=1.3) and deep in the localized phase ($W=4$). We compare them to the theoretical predictions and further contrast them with the properties of the system in the vicinity of the transition point $W_c^{}\approx 2.18-2.25$. (For illustration purposes, we chose a single value, $W=2.22$, as a representative value close to the middle of this interval.)

\subsection{Density of states and level spacing distribution} 
 \label{3D case spectral form factor}

The RMT assumes a flat density of states of a system. This assumption is strongly violated for all values of $W$ in the present model, as can be seen in Fig.$~$\ref{fig:dos_3daa}. To compare our results to RMT predictions, we therefore start with the unfolding of the spectra, which we briefly recapitulate. The spectral unfolding eliminates the influence of the density of states. It transforms an ordered set of Hamiltonian eigenvalues $\{ E_{\mu, H} \}$ into an ordered set of unfolded eigenvalues $\{ \tilde{E}_{\mu, H} \}$, for which the local mean level spacing is set to unity at all energy densities $(V^{-1}\sum_{\mu}\delta\tilde{E}_{\mu, H}= 1)$. Therefore, the scaled Heisenberg time (i.e., the inverse mean level spacing of the unfolded spectrum) is $\tilde{\tau}_H=1$. Following the standard unfolding procedure, we introduce the cumulative spectral function $\mathcal{G}(E) = \sum_{\mu}\Theta(E-E_{\mu, H})$, where $\Theta$ is the unit step function. The stepwise distribution function is then fitted by polynomials $g_n(E)$ of degree $n$ to $\mathcal{G}(E)$, and the unfolded eigenvalues are defined as $\tilde{E}_{\mu, H} = g_n(E_{\mu, H})$. We tested (not shown) the range $n=3-20$ in our calculations, and we did not observe significant changes in our results within this range of $n$; we therefore set $n=20$.

We define the gap between the two consecutive unfolded levels $\tilde{E}_{\mu,H}$ and $\tilde{E}_{\mu+1,H}$ of Hamiltonian $H$ as an analog of Eq.$~$\eqref{eq:gap_def}, i.e.,
\begin{equation}\label{eq:gap_def_un}
    \tilde{s}_{\mu,H} = \tilde{E}_{\mu+1,H}-\tilde{E}_{\mu,H}.
\end{equation}
Similarly, the gap ratio of the unfolded spectrum acts as an analog of Eq.$~$\eqref{eq:gap_ratio_def}, i.e.,
\begin{equation}\label{eq:gap_ratio_def_un}
    \tilde{r}_{\mu,H} = \frac{{\rm min}(\tilde{s}_{\mu,H}, \tilde{s}_{\mu+1,H})}{{\rm max}(\tilde{s}_{\mu,H}, \tilde{s}_{\mu+1,H})}.
\end{equation}
We note that the definition in Eq.$~$\eqref{eq:gap_ratio_def} is a bit redundant, as the gap ratio defined in Eq.$~$\eqref{eq:gap_ratio_def} already eliminates the effects of the non-flat density of states, and we do not expect differences between $P(\tilde{r}_{\mu,H})$ and $P({r}_{\mu,H})$. However, for completeness, we compare distributions of $P( \tilde{s}_{\mu,H})$ to the corresponding distributions of $P(\tilde{r}_{\mu,H})$. We collect distributions of $P( \tilde{s}_{\mu,H})$ and $P( \tilde{r}_{\mu,H} )$ from $50\%$ of the eigenstates around the mid-spectra and over $100$ Hamiltonian realizations. 

\begin{figure}[t]
  \includegraphics[width=0.48\textwidth]{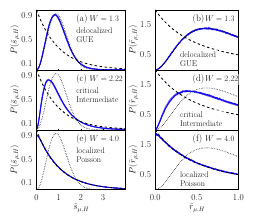}
    \caption{The  unfolded level spacing and the gap ratio distributions for the 3D Aubry-André model for $\theta=\pi/7$ and $\beta=(\sqrt{5}-1)/2$ in the closest integer approximation of the $B$-matrix. The representative values of the potential $W$ are: (a)-(b) $W=1.3$ for the delocalized phase,  (e)-(f) $W=4$ for the localized phase and,  (c)-(d) $W=2.22$ fat the estimated critical point. The dotted lines represent the theory predictions for GUE, the dashed lines represent the theory predictions for Poisson statistics. The color code denoting the systems sizes is the same as in Fig.$~$\ref{fig:r_ratio}. 
    We collect the histograms from  $50\%$ of the eigenstates around the mid-spectra and over $100$ Hamiltonian realizations. }
  \label{fig:s_3daa}
\end{figure}

The level distributions of $P( \tilde{s}_{\mu,H})$ should be compared to the theoretical RMT predictions$~$\cite{mehta_91,atas_bogomolny_13} $P_{\rm GUE}(x)=\frac{32}{\pi^2}x^2e^{-\frac{4}{\pi} x^2}$ and  $P_{\rm Pois}(x)=e^{-x}$ for the delocalized and localized phases, respectively. In Figs.$~$\ref{fig:s_3daa}(a) and$~$\ref{fig:s_3daa}(e), we confirm a good agreement between our results and the RMT predictions. Analogously, the gap ratio distributions of $P( \tilde{r}_{\mu,H} )$ should be compared to the RMT predictions$~$\cite{mehta_91,atas_bogomolny_13} $P_{\rm GUE}(x)=\frac{81\sqrt3}{2\pi}\frac{(x+x^2)^2}{(1+x+x^2)^4}$ and $P_{\rm Pois}(x)=\frac{2}{(1+x)^2}$ for the delocalized and localized phases, respectively. In Figs.$~$\ref{fig:s_3daa}(b) and$~$\ref{fig:s_3daa}(f), we again confirm a good agreement between our results and the RMT predictions.

For the representative value of the transition point $W=2.22$, we observe rather stable distributions that are neither GUE nor Poisson; see Figs.$~$\ref{fig:s_3daa}(c) and$~$\ref{fig:s_3daa}(d). This is known as intermediate statistics$~$\cite{Lozej24,jiricek2025universal}, one of the hallmarks of an eigenstate transition.

\subsection{Spectral form factor and compressibility}
 \label{3D case filter}

In this section, we go beyond the short-range spectral correlations and, with the help of the spectral form factor (SFF), study the long-range spectral correlations and compare them to the theoretical predictions. As in the case of the short-range correlations, we consider unfolded spectra. With that, we define the normalized SFF as$~$\cite{suntajs_bonca_20a}:
\begin{equation}
\tilde{K}^{}_n(\tilde{\tau}) = \frac{1}{Z}\Bigg\langle \Bigg| \sum_{\mu=1}^{V} \rho( \tilde{E}_{\mu, H}) e^{-i 2\pi\tilde{E}_{\mu, H} \tilde{\tau}} \Bigg|^2 \Bigg\rangle_H,
\label{eqn:K_uf}
\end{equation}
where $\{\tilde{E}_{\mu, H} \}$ is the complete set of Hamiltonian eigenvalues after spectral unfolding, $ \tilde{\tau}$ is the scaled time, and $Z=\langle\sum_{\mu}|\rho(\tilde{E}_{\mu, H})|^2\rangle_H$ is a normalization such that $K( \tilde{\tau} \gg1)\approx 1$. The filter $\rho(\tilde{E})$ is a function that eliminates the contributions of spectral edges. We note that each of the unfolded spectra is filtered separately.

\begin{figure}[t]
  \includegraphics[width=0.48\textwidth]{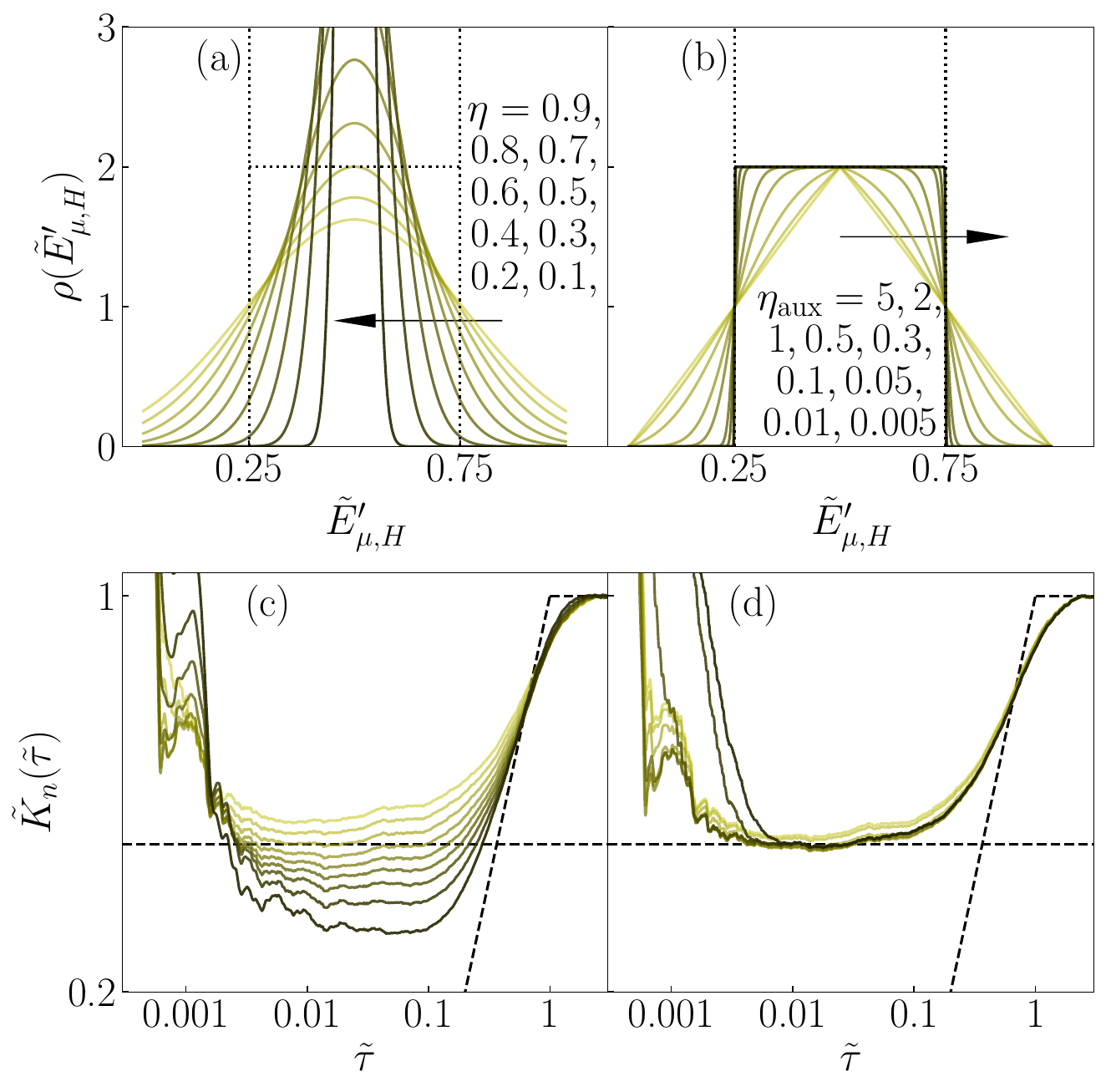}
    \caption{(a) The Gaussian and (b) the smooth box spectral filters $\rho({\tilde{E}'}_{\mu, H})$, as functions of the rescaled energy ${\tilde{E}'}_{\mu, H}\in[0,1]$. The smooth box filter is a convolution of a box function and a Gaussian function.
    The vertical dash-dotted lines correspond to the $50\%$ of energy levels around the middle of the spectrum. The horizontal dotted line represents the sharp box filter function.   
    In (c) and (d), there are examples of the normalized spectral form factor $\tilde{K}^{}_n(\tilde{\tau})$ corresponding to the filters in (a) and (b), respectively, for the 3D Aubry-André model, for $W=2.22$, $L=40$, $\theta=\pi/7$ and $\beta=(\sqrt{5}-1)/2$ in the closest determinant approximation of $B$-matrix. We average over $100$ Hamiltonian realizations.
     The horizontal dotted line represents the spectral compressibility of $50\%$ of the eigenstates around the mid-spectra, see the discussion in the main text. 
    }
  \label{fig:filter}
\end{figure}

We first discuss a Gaussian filter that was used in recent studies of ergodicity$~$\cite{suntajs_bonca_20a, Sierant2020, suntajs_prosen_21, suntajs_vidmar_22, hopjan2023,hopjan2023scale,jiricek2025universal,andreoni2026}, and we argue that it is inapplicable in the present model.
The Gaussian  filter is defined as $\rho(\tilde{E}_{\mu, H}) = \exp\{-\frac{(\tilde{E}_{\mu, H}-\bar{E}_{H})^2}{2(\eta \Gamma_{H})^2}\}$, where $\bar{E}_{H}$ and $\Gamma_{H}^2$ are the average energy and the variance, respectively, at a given Hamiltonian realization. A dimensionless parameter $\eta$ controls the effective fraction of eigenstates included in $\tilde{K}^{}_n(\tilde{\tau})$. 
In Fig.$~$\ref{fig:filter}(a), we show the Gaussian filter as a function of energy ${\tilde{E}'}_{\mu, H}\in[0,1]$ renormalized to the width of the energy band, i.e., ${\tilde{E}'}_{\mu, H}=({\tilde{E}}_{\mu, H}-{\tilde{E}}_{\mu, H}^{\rm min})/({\tilde{E}}_{\mu, H}^{\rm max}-{\tilde{E}}_{\mu, H}^{\rm min})$, for several values of the parameter $\eta = 0.1,\dots ,0.9$ in steps of $0.1$. We observe that as $\eta$ increases, we include more states at the edge of the spectrum. In Fig.$~$\ref{fig:filter}(c), we show the corresponding normalized SFF from Eq.$~$\eqref{eqn:K_uf}, which results from the Gaussian filters in Fig.$~$\ref{fig:filter}(a). Unlike in standard systems, see e.g., $~$\cite{suntajs_bonca_20a,jiricek2025universal}, we observe a strong dependence of the SFF on the parameter $\eta$. This behavior is another manifestation of the non-smooth structures of the level spacings, see Fig.$~$\ref{fig:r_ratio}, induced by the commensurate approximations. Thus, we should avoid using the Gaussian filter in the present model.

To resolve this issue, instead of the Gaussian filter, we design a filter that reflects the specific choice of averaging over $50\%$ of the eigenstates around the mid-spectra, as discussed in Sec.$~$\ref{3D case edges}. The natural choice is a box filter function that includes exactly $50\%$ of the eigenstates around the mid-spectra. We show such a function in Fig.$~$\ref{fig:filter}(b). However, the sharp structure of the box filter function can induce unwanted effects in the spectral form factor. To smooth the box filter, we first convolve it with an auxiliary Gaussian function $\rho_{\rm aux}(\tilde{E}_{\mu, H}) = \exp\{-\frac{(\tilde{E}_{\mu, H}-\bar{E}_{H})^2}{2(\eta_{\rm aux} \Gamma_{H})^2}\}$ with a width controlled by a parameter $\eta_{\rm aux}$. As both the box and Gaussian functions are defined on a finite interval $[0,V]$, the convolution is given for $[0,2V]$. To obtain the smooth box function, we squeeze the convolution back to the interval $[0,V]$. 

Examples of the smooth box filter functions as a function of the renormalized energy ${\tilde{E}'}_{\mu, H}\in[0,1]$ are shown in Fig.$~$\ref{fig:filter}(b) for several values of the parameter $\eta_{\rm aux} \in \{5,2,1,0.5,0.3,0.1,0.05,0.01,0.005\}$. We observe that, with increasing width of the auxiliary Gaussian $\eta_{\rm aux}$, the box function becomes increasingly smooth at points $1/4$ and $3/4$, ultimately converging to the triangle function. The triangle function represents the convolution of two box functions, one being the original box function and the second being the Gaussian with infinite width, numerically represented in a finite interval as a box function. In Fig.$~$\ref{fig:filter}(d), we plot the corresponding normalized SFF. Importantly, contrary to the case of the Gaussian filter function, we observe that the normalized SFF is independent of $\eta_{\rm aux}$ over a broad interval of times when $\eta_{\rm aux}$ is not too large, i.e., $\eta_{\rm aux}<1$. We choose $\eta_{\rm aux}=0.1$ as the optimal value of $\eta_{\rm aux}$.

\begin{figure}[t]
  \includegraphics[width=0.48\textwidth]{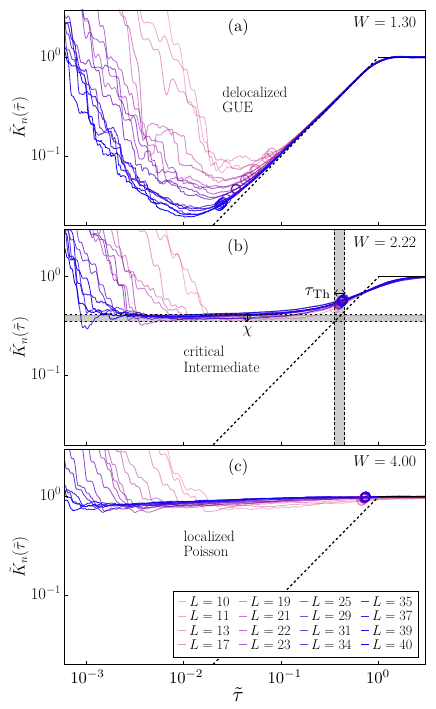}
  \caption{The normalized SFF $\tilde{K}^{}_n(\tilde{\tau})$ at (a) $W = 1.3$ (delocalized), (b) $W = 2.22$ (critical) and (c) $W = 4.0$ (localized) for the 3D Aubry-André model for $\theta=\pi/7$ and $\beta=(\sqrt{5}-1)/2$ in the closest integer approximation of the $B$-matrix. We use $n=20$ and $\eta_{\rm aux}=0.1$ and average over $100$ Hamiltonian realizations. We consider the moving averages in time over $300$ consecutive discrete time points. The dashed lines denote the GUE prediction $\tilde{K}^{{\rm GUE}}_{n}(\tilde{\tau}) = \tilde{\tau}$. The empty circles represent the Thouless times.
  The horizontal and vertical shaded region mark the range of compressibility and the Thouless time fluctuations, respectively. }  
  \label{fig:sff_unfold}
\end{figure}

Having chosen a suitable filter function, we can study the behavior of the normalized SFF $\tilde{K}^{}_n(\tilde{\tau})$ across the delocalization-localization transition. We compare our results with the infinite system size predictions, i.e., the GUE predictions $\tilde{K}^{{\rm GUE}}_{n}(\tilde{\tau}) = \tilde{\tau}$ and the Poisson statistics prediction $\tilde{K}^{{\rm Pois}}_{n}(\tilde{\tau})= 1$, which are valid for $\tilde{\tau}<1$. 

The onset of universal behavior in a finite system occurs at the Thouless time $\tilde{\tau}_{{\rm Th}}$. To numerically obtain its value, we first calculate $\tilde{K}^{}_n(\tilde{\tau})$ for $N_{\tilde{\tau}}=5000$ discrete time steps $\tilde{\tau}_i$ on the logarithmic grid. We then smooth out fluctuations in $\tilde{K}^{}_n(\tilde{\tau})$ by calculating a running mean over $300$ values of $K_n(\tilde{\tau}_i)$. Finally, we interpolate the smooth normalized SFF with $10000$ points. With the interpolated points, we then analyze the difference between $\tilde{K}^{}_n(\tilde{\tau})$ and $\tilde{K}^{{\rm GUE}}_{n}(\tilde{\tau}) = \tilde{\tau}$, using a deviation measure 
\begin{equation}
    \Delta \tilde{K}(\tilde{\tau}) = \Bigg|\ln\Bigg(\frac{\tilde{K}^{}_n(\tilde{\tau})}{\tilde{K}^{{\rm GUE}}_{n}(\tilde{\tau})}\Bigg)\Bigg|;\hspace{3mm} \Delta \tilde{K}(\tilde{\tau}_{{\rm Th}}) = \epsilon. 
    \label{eqn:K_dev} 
\end{equation}
Here, the scaled Thouless time $\tilde{\tau}_{{\rm Th}}$ is the time at which $\Delta K(\tilde{\tau})$ becomes greater than a chosen threshold value $\epsilon$. We note that although the absolute value of the extracted Thouless time depends on the threshold $\epsilon$, a power-law exponent characterizing the system size dependence of the Thouless time is expected to be independent of this parameter. We have checked (not shown) that the exponent of the Thouless time system-size dependence does not change significantly for $\epsilon=0.1-0.9$. Here, we choose $\epsilon=0.3$.

In Fig.$~$\ref{fig:sff_unfold}(a), we show the smooth normalized SFF in the delocalized phase at $W=1.3$ for different system sizes $L$. We observe that the finite-size curves of $\tilde{K}^{}_n(\tilde{\tau})$ do approach the universal GUE ramp $\tilde{K}^{{\rm GUE}}_{n}(\tilde{\tau})$ for $\tilde{\tau}>\tilde{\tau}_{{\rm Th}}$. We observe that the Thouless time $\tilde{\tau}_{{\rm Th}}$ decreases with increasing system size $L$; see the empty circles in Fig.$~$\ref{fig:sff_unfold}(a). 
Analogously, in Fig.$~$\ref{fig:sff_unfold}(c), we show the smooth normalized SFF in the localized phase at $W=4$ for different system sizes $L$.  We observe that the finite-size curves of $\tilde{K}^{}_n(\tilde{\tau})$ do approach the Poisson prediction $\tilde{K}^{{\rm Pois}}_{n}(\tilde{\tau})= 1$ and that $\tilde{\tau}_{{\rm Th}}$ slightly increases with the system size. 

For the representative value of the transition point $W=2.22$, see Fig.$~$\ref{fig:sff_unfold}(b). The smooth normalized SFF initially decreases, then reaches a constant value for a broad mid-time interval, and finally goes to unity at late times. The value of $\tilde{K}^{}_n(\tilde{\tau})$ of the mid-time plateau is known as the spectral compressibility $\chi$$~$\cite{Chalker96, Bogomolny2011,jiricek2025universal}. The system-size independent plateau value $\chi\propto {\rm const.}$, together with the system independent Thouless time $\tilde{\tau}_{{\rm Th}}\propto {\rm const.}$, are hallmarks of an eigenstate transition$~$\cite{suntajs_bonca_20a, Sierant2020, suntajs_prosen_21, suntajs_vidmar_22, hopjan2023,hopjan2023scale,jiricek2025universal}. Figure$~$\ref{fig:sff_unfold}(b) suggests that these hallmarks are consistently present for the representative value of the transition point $W=2.22$.

That being said, we notice that even though the plateaus in Fig.$~$\ref{fig:sff_unfold}(b) are well developed, the values of the spectral compressibility exhibit slight fluctuations; see the shaded horizontal area in Fig.$~$\ref{fig:sff_unfold}(b). The same is true for the value of the Thouless time $\tilde{\tau}_{{\rm Th}}\propto {\rm const.}$; see the shaded vertical area in Fig.$~$\ref{fig:sff_unfold}(b). The fluctuations are induced by the imposition of the commensurate approximations of the $B$-matrix and reflect the fact that the approximation is slightly different for each system size. {In Appendix$~$\ref{3D case fluctspec}, we demonstrate that the fluctuations are consistently present in all spectral measures studied here.

\subsection{Testing universal relation between spectral and wave-function properties at criticality}
 \label{3D case universal}
 
It was conjectured$~$\cite{Chalker96, Bogomolny2011} and confirmed for several models$~$\cite{jiricek2025universal} that there exists a universal relation between the spectral and wave-function properties at a critical point. More specifically, there exists a simple connection between the spectral compressibility $\chi$ and the $q=1$ fractal dimension $d^{pb,(1)}_{{\rm typ}}$ in a preferred basis $|pb\rangle$, where $q$ represents the power of the moments of wave-function amplitudes. This relation can be expressed as $\chi+d^{pb,(1)}_{{\rm typ}}=1$. The relation was tested in various quadratic lattice models at their transition points$~$\cite{jiricek2025universal}. To test the conjecture on our results for the 3D Aubry-André model, we determine the values of $\chi$ and $d^{l,(1)}_{{\rm typ}}$ below, assuming that the preferred basis is the lattice basis. We consider both commensurate approximations from Sec.$~$\ref{choice dual approx}, i.e., the closest integer and the closest determinant approximation, and we study whether our conclusions depend on the particular approximation.

\begin{figure}[t]
\includegraphics[width=0.45\textwidth]{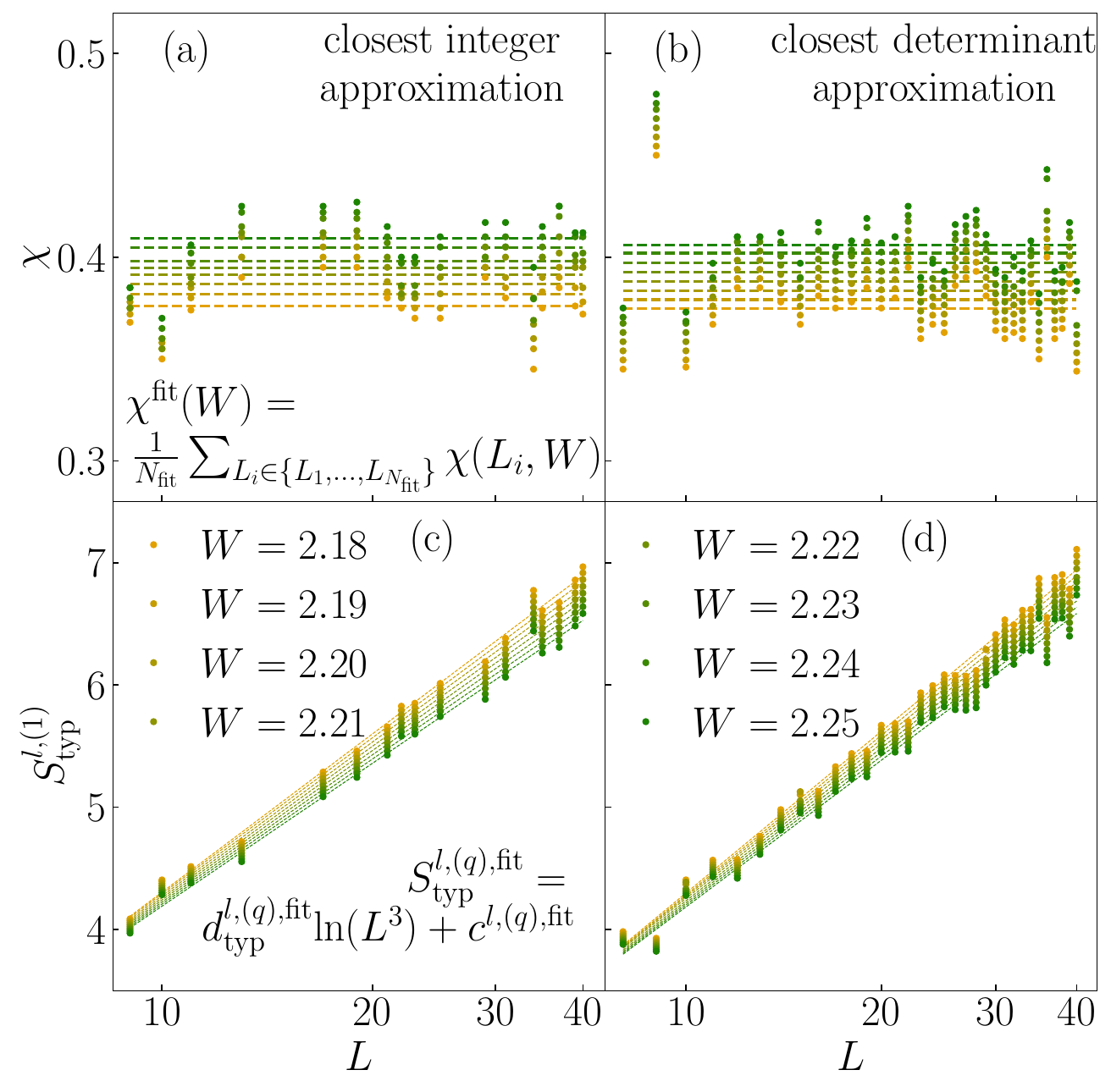}
  \caption{{The spectral and wave-function properties of the 3D Aubry-André model for $\theta=\pi/7$ and $\beta=(\sqrt{5}-1)/2$ in (a), (c) the closest integer approximation of the $B$-matrix, and (b), (d) the closest determinant approximation of $B$-matrix. 
  (a)-(b) Finite-size dependence the compressibility $\chi$. It is extracted from the normalized SFF with the smooth box function with $\eta_{\rm aux}=0.1$. We average over $100$ Hamiltonian realizations. The dashed lines in panels (a)-(b) represent examples of fit, Eq.$~$\eqref{eqn:chi_fit}, of the compressibility $\chi$ for several values of the potential strength $W$.
  (c)-(d) The participation entropy $S_{\rm typ}^{l,(1)}$ for $q=1$ and for several values of the potential strength $W$. The dashed lines
  represent examples of fit in Eq.$~$\eqref{eqn:Sq_fit}.
  We average over $50\%$ eigenstates around the mid spectra and $100$ Hamiltonian realizations. }
  }
  \label{fig:Dq_typ}
\end{figure}

We first discuss the compressibility. It was observed in Fig.$~$\ref{fig:sff_unfold}(b) that the value of compressibility $\chi$ fluctuates with the system size. We illustrate this further in Figs.$~$\ref{fig:Dq_typ}(a) and$~$\ref{fig:Dq_typ}(b), where the values of $\chi$ are plotted as a function of system size $L$ for $W=2.18-2.25$ in steps of $0.01$. Indeed, the fluctuations are present. However, in Appendix$~$\ref{3D case fluctspec}, we demonstrate that they are consistently present in all other spectral measures studied here. This fact allows us to separate the fluctuations from the main trend. At the transition point, the compressibility is assumed to be constant as the main trend. We thus fit the values of $\chi(L)$ in Figs.$~$\ref{fig:Dq_typ}(a) and$~$\ref{fig:Dq_typ}(b) with a constant function, which is equivalent to finding the mean of $\chi(L)$, i.e., 
\begin{equation}
\chi^{{\rm fit}}(W)=\frac{1}{N_{\rm fit}} \sum_{L_i\in\{L_1,\dots,L_{N_{\rm fit}}\}}\chi(L_i,W),
\label{eqn:chi_fit}
\end{equation}
where $N_{\rm fit}$ is the number of values available for each $W$. The results of the fits are shown in Figs.$~$\ref{fig:Dq_typ}(a) and$~$\ref{fig:Dq_typ}(b) as dashed lines. The values of $\chi^{{\rm fit}}$ from Figs.$~$\ref{fig:Dq_typ}(a) and$~$\ref{fig:Dq_typ}(b) are reported in Appendix$~$\ref{3D case fluctwave} in Figs.$~$\ref{fig:fluct}(a) and$~$\ref{fig:fluct}(b), respectively.

The values of compressibility should be compared to the values of the fractal dimension. The fractal dimension can be extracted either from the scaling of the average IPR, see Sec.$~$\ref{3D case scaling}, or from the scaling of the average participation entropy. Roughly speaking, the average participation entropy provides information about typical values of wave-function amplitudes rather than average ones. The conjecture of the universal relation between spectral and wave-function properties at a critical point was successfully tested for the typical fractal dimension extracted from the participation entropies$~$\cite{jiricek2025universal}, and thus we here also consider the typical fractal dimension. In Appendix$~$\ref{3D case fluctwave}, we provide corresponding definitions of the participation entropies for general $q$ moments of wave-function amplitudes, i.e., different from $q=2$ as in the IPR.

For illustration, in Figs.$~$\ref{fig:Dq_typ}(c) and$~$\ref{fig:Dq_typ}(d), we plot the $q=1$ participation entropy $S^{l,(1)}_{{\rm typ}}$ for several values of the potential strength $W$ close to the eigenstate transition. As the main trend, we observe logarithmic growth of the participation entropy with $L$. Similar to the spectral measures, we observe additional fluctuations on top of the main trend. In Appendix$~$\ref{3D case fluctwave}, we demonstrate that the fluctuations in the participation entropies are similar to the fluctuations in spectral measures, and thus can ultimately be mitigated.

In order to separate the fluctuations from the main trend, we fit the participation entropy with an ansatz
\begin{equation}
    S^{l,(q),{\rm fit}}_{{\rm typ}} = d^{l,(q),{\rm fit}}_{{\rm typ}}\ln (L^3) + c^{l,(q),{\rm fit}},
    \label{eqn:Sq_fit} 
\end{equation}
where the slope $d^{l,(q),{\rm fit}}_{{\rm typ}}$ of the fit can be interpreted as the fractal dimension of the $q$-moment of the wave-function. For a perfectly delocalized state, $d^{l,(q),{\rm fit}}_{{\rm typ}}=1$ for all $q$. On the other hand, for a localized state, $d^{l,(q),{\rm fit}}_{{\rm typ}}=0$. In an intermediate case, $0<d^{l,(q),{\rm fit}}_{{\rm typ}}<1$, the state is fractal if $d^{l,(q),{\rm fit}}_{{\rm typ}}$ is the same for all $q$, or multifractal$~$\cite{mace2019multifractal, lindinger2019many} if $d^{l,(q),{\rm fit}}_{{\rm typ}}$ is not the same for all $q$. 

Examples of the fits of $S^{l,(1)}_{{\rm typ}}$ are plotted as dashed lines in Figs.$~$\ref{fig:Dq_typ}(c) and$~$\ref{fig:Dq_typ}(d). For other $q$-values the picture is similar (not shown). The values of $d^{l,(1),{\rm fit}}_{{\rm typ}}$ from Figs.$~$\ref{fig:Dq_typ}(c) and$~$\ref{fig:Dq_typ}(d) are reported in Appendix$~$\ref{3D case fluctwave} in Figs.$~$\ref{fig:fluct}(c) and$~$\ref{fig:fluct}(d), respectively.}

\begin{figure}[t]
\includegraphics[width=0.45\textwidth]{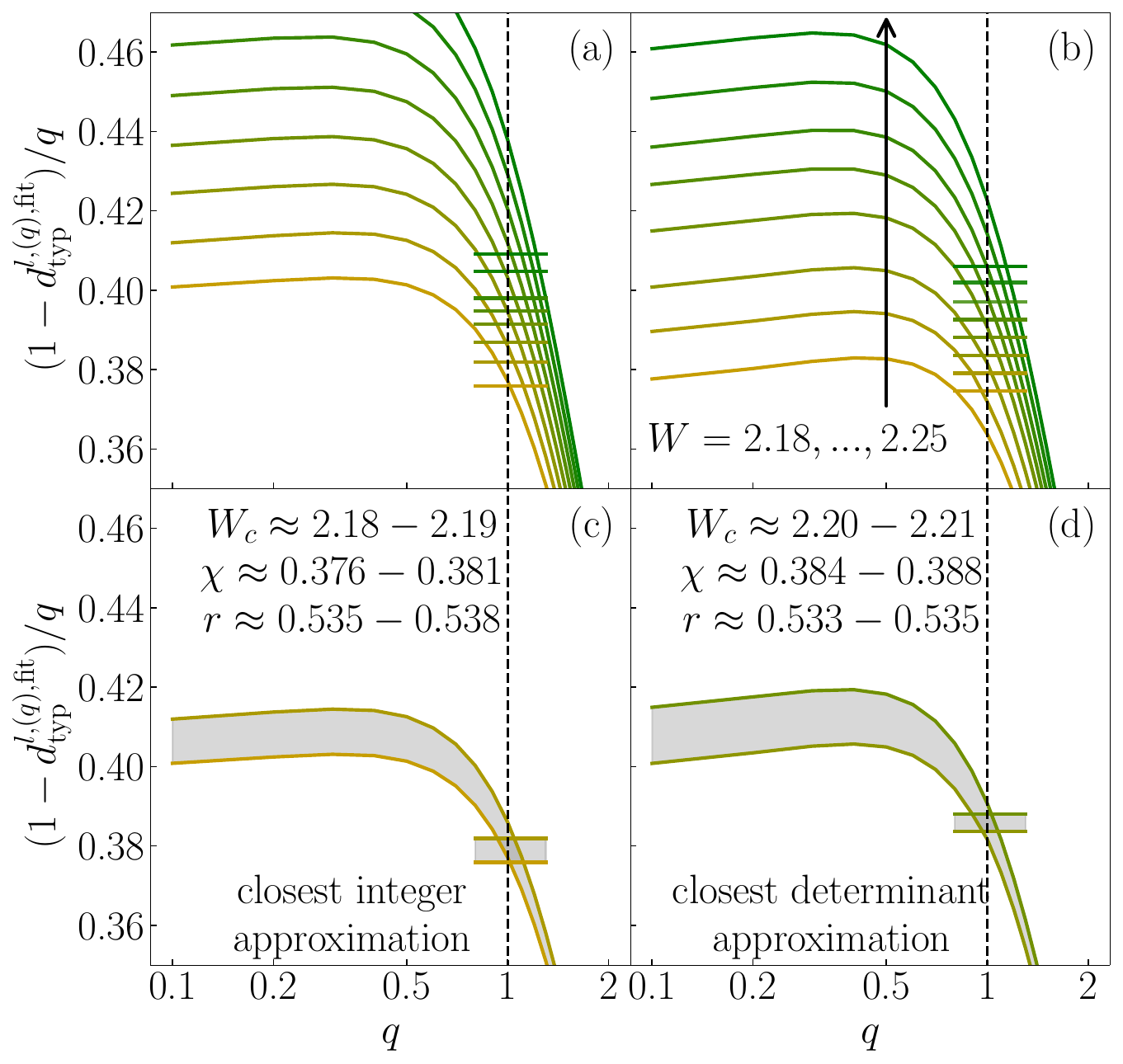}
  \caption{Testing the universal relation between spectral and wave-function properties at criticality in Eq.$~$\eqref{eqn:relation} in (a), (c) the closest integer approximation of the $B$-matrix, and (b), (d) the closest determinant approximation of $B$-matrix.
  (a)-(b) The rescaled fractal dimension 
  $\big(1-d^{l,(q),{\rm fit}}_{{\rm typ}}\big)/q$ as a function of $q$ for $W=2.18-2.25$ in steps of $0.01$. Horizontal lines are corresponding values of the compressibility $\chi^{{\rm fit}}$. (c)-(d)  The rescaled fractal dimension 
  $\big(1-d^{l,(q),{\rm fit}}_{{\rm typ}}\big)/q$
 for range of $W$ where the relation in Eq.$~$\eqref{eqn:relation} is fulfilled, i.e., the estimated transition point.
  }
  \label{fig:chi_Dq_typ}
\end{figure}

We now test the conjecture on our results for the 3D Aubry-André model. Since the fluctuations are similar in both the spectral and wave-function measures, to test the conjecture, we use the quantities that describe the main trends, i.e., $\chi^{{\rm fit}}$ and $d^{l,(1),{\rm fit}}_{{\rm typ}}$. Then, the universal relation can be adapted to
\begin{equation}
    \chi^{{\rm fit}}+d^{l,(1),{\rm fit}}_{{\rm typ}}=1.
    \label{eqn:relation} 
\end{equation}
In the vicinity of $W_c^{}$, we choose the preferred basis to be the lattice basis, i.e., $|pb\rangle=|l\rangle$, as the preferred basis is usually the one in which the model localizes$~$\cite{jiricek2025universal}. In the vicinity of $1/W_c^{}$, the preferred basis is the dual momentum basis, i.e., $|pb\rangle=|k^{\rm D}\rangle$, as the localization occurs there.

In Figs.$~$\ref{fig:chi_Dq_typ}(a) and$~$\ref{fig:chi_Dq_typ}(b), we show the behavior of $(1-d^{l,(q),{\rm fit}}_{{\rm typ}})/q$ as a function of $q$. For $q=1$, denoted as the vertical dot-dashed line, this reduces to $1-d^{l,(q),{\rm fit}}_{{\rm typ}}$, and it should be compared to the compressibility values $\chi^{{\rm fit}}$, denoted as the horizontal dashed lines. In Figs.$~$\ref{fig:chi_Dq_typ}(c) and$~$\ref{fig:chi_Dq_typ}(d), we highlight the curves from Figs.$~$\ref{fig:chi_Dq_typ}(a) and$~$\ref{fig:chi_Dq_typ}(b), respectively, for values $W$ for which Eq.$~$\eqref{eqn:relation} is satisfied to high precision. We report the particular values of $W_c^{}$, $\chi^{{\rm fit}}$, and $r$ at the transition point. Remarkably, the values of the transition point are very similar; the transition point is found at $W_c^{}\approx2.18-2.19$ and $W_c^{}\approx2.20-2.21$ in the closest integer approximation and the closest determinant approximation, respectively.

The value of the transition point from the test of the conjecture in Eq.$~$\eqref{eqn:relation}, i.e., $W_c^{}\approx2.18-2.21$, should be compared to the value from the scaling analysis in Sec.$~$\ref{3D case scaling}, i.e., $W_c^{}\approx2.19-2.24$. They differ by only one percent, which is a sign of the consistency of our analysis. Naturally, we can ask which analysis is better. The value from the scaling analysis in Sec.$~$\ref{3D case scaling} is based on the raw curves, i.e., the curves containing the fluctuations, and the value determined in this section is based on the curves from which the fluctuations were removed. Thus, we believe that the values in this section are more accurate for the available system sizes. 

This concludes our study of the static properties of the present model, and in the next sections, we study implications for the dynamics.

\section{Wave-packet transport and critical dynamical exponent at eigenstate transition} \label{3D case time}

In Sec.$~$\ref{3D case} we studied the static properties of the 3D Aubry-André model and show that the eigenstate transition belongs to the class of ergodic-to-localized phase transitions. The transition point at $W_c\approx2.18-2.21$, which is associated with real-space localization, is similar to the one in the paradigmatic 3D Anderson model. It is known that transport at the critical point of the 3D Anderson model is sub-diffusive$~$\cite{Ohtsuki97,Sierant2020} which is the consequence of the criticality$~$\cite{hopjan2025critical}. Given the similarity of the static properties between the 3D Aubry-André model and the 3D Anderson model at the transition point, in this section, we ask whether analogous similarity applies to the dynamical properties, i.e., whether transport at the critical point is also sub-diffusive. Before answering this question, we first look at the basic dual symmetry of the dynamical properties of the 3D Aubry-André model, valid for any $W$.

{
\subsection{Symmetry of the transport} \label{3D case time overlap}

In Secs.$~$\ref{3D case duality spectra} and$~$\ref{3D case eigenstates}, we have argued for the symmetries in the eigenspectra and the eigenvectors, respectively, and we have concluded that properties in real-space at $W$ are symmetric to the properties in the dual momentum space at $1/W$. Thus, here we ask whether we can connect transport in real space at $W$ to the transport in dual momentum space at $1/W$.

To do so, we consider the following initial conditions,
\begin{align}\nonumber
&|l_0\rangle=c_{\vec{l}_0}^{\dagger}|\varnothing\rangle,\\\label{eq:initial0}
&|\tilde{k}_0^{\rm D}\rangle=e^{i2\vec{k}^{\rm D}\vec{\phi}}L^{-3/2}\sum_{\vec{l}}e^{i 2\pi \vec{k}_0^{\rm D}\cdot B^{\rm CA} \cdot\vec{l}}c^{\dagger}_{\vec{l}}|\varnothing\rangle,
\end{align}
where $\tilde{k}_0^{\rm D}$ was created by the unitary transforms described in Eqs.$~$\eqref{eqn:H_3DAA_ftsd} and$~$\eqref{eqn:H_3DAA_phase_twist}. 

We compare the transport of a particle in real space under the evolution of the real space Hamiltonian at $W$, to the transport of a particle in dual space under the evolution of the real space Hamiltonian at $1/W$. As a first step, we use the unitary transform in Eq.$~$\eqref{eqn:dual_trans_twist} to rewrite the overlap in momentum space evolved by the real-space Hamiltonian at $1/W$, to the following form:
\begin{align}
\nonumber
&\langle \tilde{k}^{\rm D}|e^{-i{H_{\rm CA}^{l,\phi}[1,2/W] }t}|\tilde{k}_0^{\rm D}\rangle = \langle \tilde{k}^{\rm D}|e^{-i{\tilde{H}_{\rm CA}^{k,\phi}[1/W,2] }t} |\tilde{k}_0^{\rm D}\rangle  = \\\label{eqn:overlaps_approx_twist}
&\langle \tilde{k}^{\rm D}|e^{-i{\tilde{H}_{\rm CA}^{k,\phi}[1,2W] }t/W} |\tilde{k}_0^{\rm D}\rangle=\langle \tilde{k}^{\rm D}|e^{-i{\tilde{H}_{\rm CA}^{k,\phi}[1,2W] } \tilde{t}} |\tilde{k}_0^{\rm D}\rangle,
\end{align}
where, in the second and third step, we extracted the factor $1/W$ from the Hamiltonian and redefined the time as
$\tilde{t}=t/W$. The last term from Eq.$~$\eqref{eqn:overlaps_approx_twist} is, up to the boundary terms, equal to the overlap in the real-space lattice,
\begin{equation}\label{eqn:overlaps_approx_twist2}
\langle \tilde{k}^{\rm D}|e^{-i{\tilde{H}_{\rm CA}^{k,\phi}[1,2W] } \tilde{t}} |\tilde{k}_0^{\rm D}\rangle  \approx 
\langle l|e^{-i{H_{\rm CA}^{l,\phi}[1,2W] }t}|l_0\rangle,
\end{equation}
as $H_{\rm CA}^{l,\phi}[1,2W]$ and $\tilde{H}_{\rm CA}^{k,\phi}[1,2W]$ have, up to the boundary terms, the same form in the $|l_0\rangle $ and $|\tilde{k}_0^{\rm D}\rangle$ bases, respectively.

Finally, from Eqs.$~$\eqref{eqn:overlaps_approx_twist} and$~$\eqref{eqn:overlaps_approx_twist2}, we obtain
\begin{equation}\label{eqn:overlaps_approx_twist3}
\langle \tilde{k}^{\rm D}|e^{-i{H_{\rm CA}^{l,\phi}[1,2/W] }t}|\tilde{k}_0^{\rm D}\rangle \approx \langle l|e^{-i{H_{\rm CA}^{l,\phi}[1,2W] }t}|l_0\rangle. 
\end{equation}
The transport of the particles in real space under the evolution of the real space Hamiltonian at $W$, and the transport of the particles in dual space under the evolution of the real space Hamiltonian at $1/W$, are equal, up to the effects of the twisted boundaries. In the thermodynamic limit, the effect of boundaries is expected to vanish, and the approximate equality should become the true equality.

Having discussed the transition probabilities, we turn to quantifying the spread of the particle by the mean square displacement, defined as 
\begin{align}\nonumber
&\sigma^2_{H,l_0}(t) = \sum_{l}D_{l,l_0}^{2} P_{l,l_0},\\\label{eqn:msd}
&\sigma^2_{H,\tilde{k}_0^{\rm D}}(t) = \sum_{\tilde{k}^{\rm D}}D_{\tilde{k}^{\rm D},\tilde{k}_0^{\rm D}}^{2} P_{\tilde{k}^{\rm D},\tilde{k}_0^{\rm D}},
\end{align}
in the real-space and dual momentum lattice, respectively. Here,
\begin{align}\nonumber
&P_{l,l_0}=|\langle l|e^{-i{H_{\rm CA}^{l,\phi}[1,2W] }t}|l_0\rangle|^2,\\\label{eqn:transitions}
&P_{\tilde{k}^{\rm D},\tilde{k}_0^{\rm D}}=|\langle \tilde{k}^{\rm D}|e^{-i{H_{\rm CA}^{l,\phi}[1,2/W] }t}|\tilde{k}_0^{\rm D}\rangle|^2,
\end{align}
are the transition probabilities$~$\cite{Jiricek24,hopjan2024survival,hopjan2025critical}. 
Further, the distances $D_{l,l_0}$ and $D_{\tilde{k}^{\rm D},\tilde{k}_0^{\rm D}}$ are the Euclidean distances between the initial site $\vec{l}_0$ and sites $\vec{l}$, and the initial site $\vec{\tilde{k}}_0^{\rm D}$ and sites $\vec{\tilde{k}}^{\rm D}$ in the real-space and dual momentum lattice, respectively. As the model is defined with periodic boundary conditions, i.e., in the commensurate approximation, the distance must reflect these conditions. Thus, the distance is defined as the minimal distance, i.e.,
\begin{align}\nonumber
&D^x_{l,l_0}={\rm min}(|l_x-l_{0,x}|,|l_x-l_{0,x}+L|,|l_x-l_{0,x}-L|),\\\nonumber
&D^y_{l,l_0}={\rm min}(|l_y-l_{0,y}|,|l_y-l_{0,y}+L|,|l_y-l_{0,y}-L|),\\\nonumber
&D^z_{l,l_0}={\rm min}(|l_z-l_{0,z}|,|l_z-l_{0,z}+L|,|l_z-l_{0,z}-L|),\\\label{eqn:distance}
&D_{l,l_0}=\sqrt{(D^x_{l,l_0})^2+(D^y_{l,l_0})^2+(D^z_{l,l_0})^2}.
\end{align}
The distance in the dual momentum lattice is defined analogously with the replacement $l \rightarrow \tilde{k}^{\rm D}$ and $l_0 \rightarrow \tilde{k}^{\rm D}_0$.

Finally, we define averages over the initial states and the Hamiltonian realizations,
\begin{align}\nonumber
&\sigma^2_{l}(t) =\Big<\Big< \sigma^2_{H,l_0}(t)  \Big>_{l_0^{}}\Big>_{H},\\\label{eqn:msd}
&\sigma^2_{\tilde{k}}(t) =\Big<\Big< \sigma^2_{H,\tilde{k}_0^{\rm D}}(t) \Big>_{\tilde{k}_0^{\rm D}}\Big>_{H}.
\end{align}
Based on Eq.$~$\eqref{eqn:overlaps_approx_twist3}, we expect that 
\begin{align}\label{eqn:msd_approx}
&\sigma^2_{H,l_0}(t)\approx \sigma^2_{H,\tilde{k}_0^{\rm D}}(t)~~~~~~~~\sigma^2_{l}(t) \approx \sigma^2_{\tilde{k}^{\rm D}}(t),
\end{align}
when considering the same starting points of the $|l_0\rangle$ and $|\tilde{k}_0^{\rm D}\rangle$ in the real-space and dual momentum lattice, respectively. 
\begin{figure}[t]
  \includegraphics[width=0.45\textwidth]{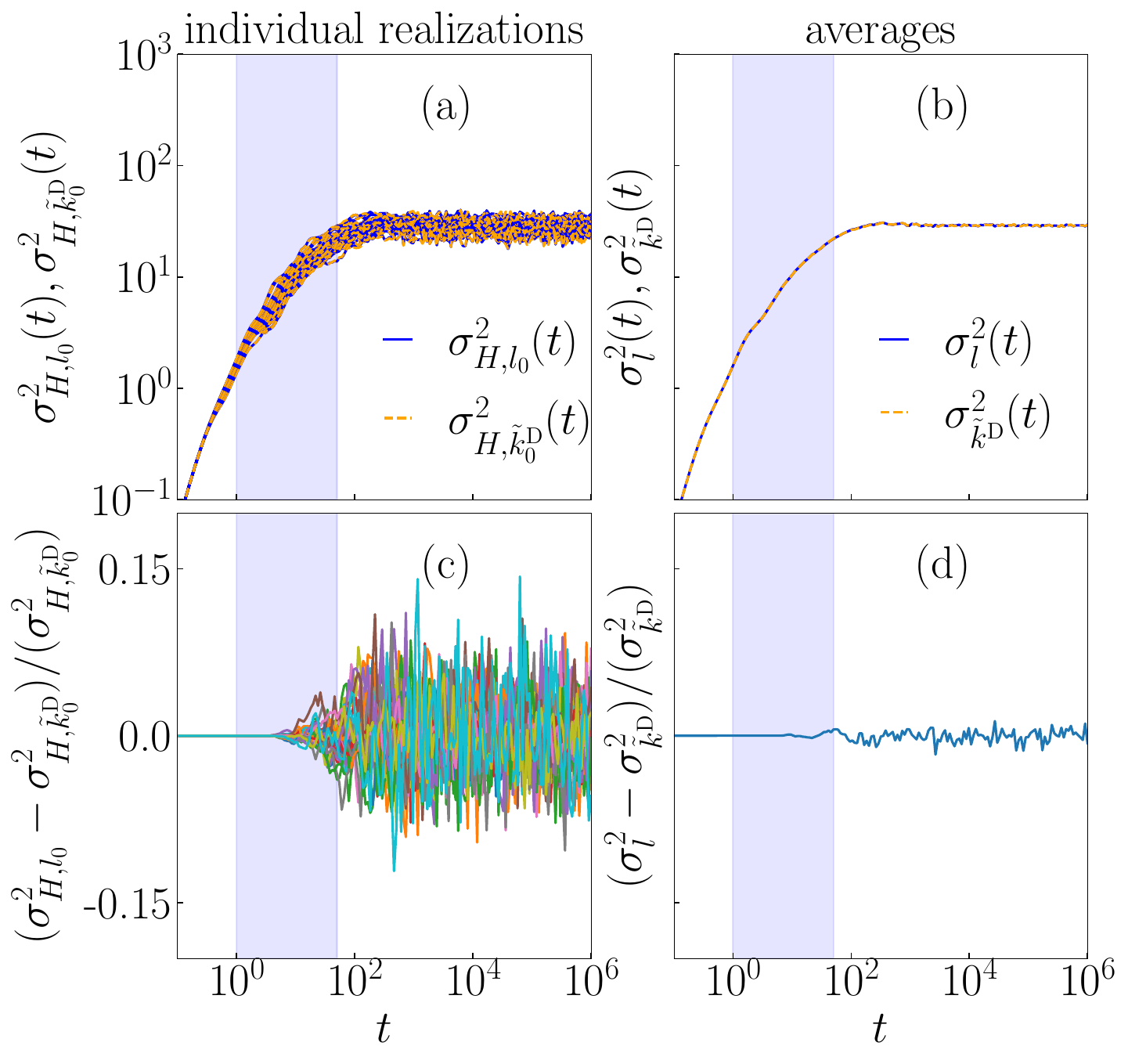}
  \caption{Comparison of the spreading of the particles in real space under evolution of the real-space Hamitonian at $W$, to the spreading of the particles in dual space under evolution of the real-space Hamitonian at $1/W$. The mean square displacement for different Hamiltonian realizations in the 3D Aubry-André model with $L=13$, $\theta=\pi/7$ and $\beta=(\sqrt{5}-1)/2$ in the closest integer approximation. The particles are initialized in the center of corresponding cubes. (a) Individual comparisons considering $30$ Hamiltonian realizations.  (b) The average of the individual realizations shown in (a). (c) The relative differences of curves in (a). (d)  The relative differences of curves in (b). The shaded area highlights the hydrodynamical regime $t_{Th}>t>1$.
  }
  \label{fig:msd_central}
\end{figure}

In Fig.$~$\ref{fig:msd_central}, we test this approximate equivalence of Eq.$~$\eqref{eqn:msd_approx} on single particles initialized in the center of the lattice. For this initial state, the best approximate equivalence is expected as the effects of the twisted boundaries are minimized. In Fig.$~$\ref{fig:msd_central}(a), we show the comparison for several Hamiltonian realizations. We observe that there is indeed qualitative agreement already at the level of individual particles, even for quite moderate system sizes. To quantify this agreement, we plot the relative error between $\sigma^2_{H,l_0}(t)$ and $\sigma^2_{H,\tilde{k}_0^{\rm D}}(t)$ in Fig.$~$\ref{fig:msd_central}(c). We observe that some differences are present. However, the larger differences appear only after the Thouless time is reached. The mid-time hydrodynamical regime, which is of interest here, exhibits only moderate differences. The situation is improved enormously for the mean square displacement averaged over Hamiltonian realizations, see Figs.$~$\ref{fig:msd_central}(b) and$~$\ref{fig:msd_central}(d). 

We can conclude that the transport in real space at $W$ is symmetric to the transport in dual momentum space at $1/W$.
Therefore, in the next sections, we study the properties of the transport in real space only, denoted as $\sigma^2 \equiv \sigma^2_{l}$, and we assume that the properties of the transport in the dual momentum space can be deduced from the restored symmetry in the thermodynamic limit.

\subsection{Scaling of mean square displacement}
\label{3D case time scaling}

We study the system size scaling of the mean square displacement. We start our discussion with Figs.$~$\ref{fig:msd}(a) and$~$\ref{fig:msd}(b) where we compare the system size scaling of the mean square displacement for the full averaging to the averaging that avoids the effect of the mobility edges discussed in Appendix$~$\ref{3D case time initial state}.} In both cases, the mean square displacement grows ballistically for a short time, after which a hydrodynamic slope develops in the mid-time regime before saturation is reached at the Thouless time. We observe that the slope of the mean square displacement is steeper when considering the selective averaging. This confirms the picture that the selected initial states, which project mainly to the critical eigenstates, delocalize more during the time evolution. 

\begin{figure}[t]
  \includegraphics[width=0.45\textwidth]{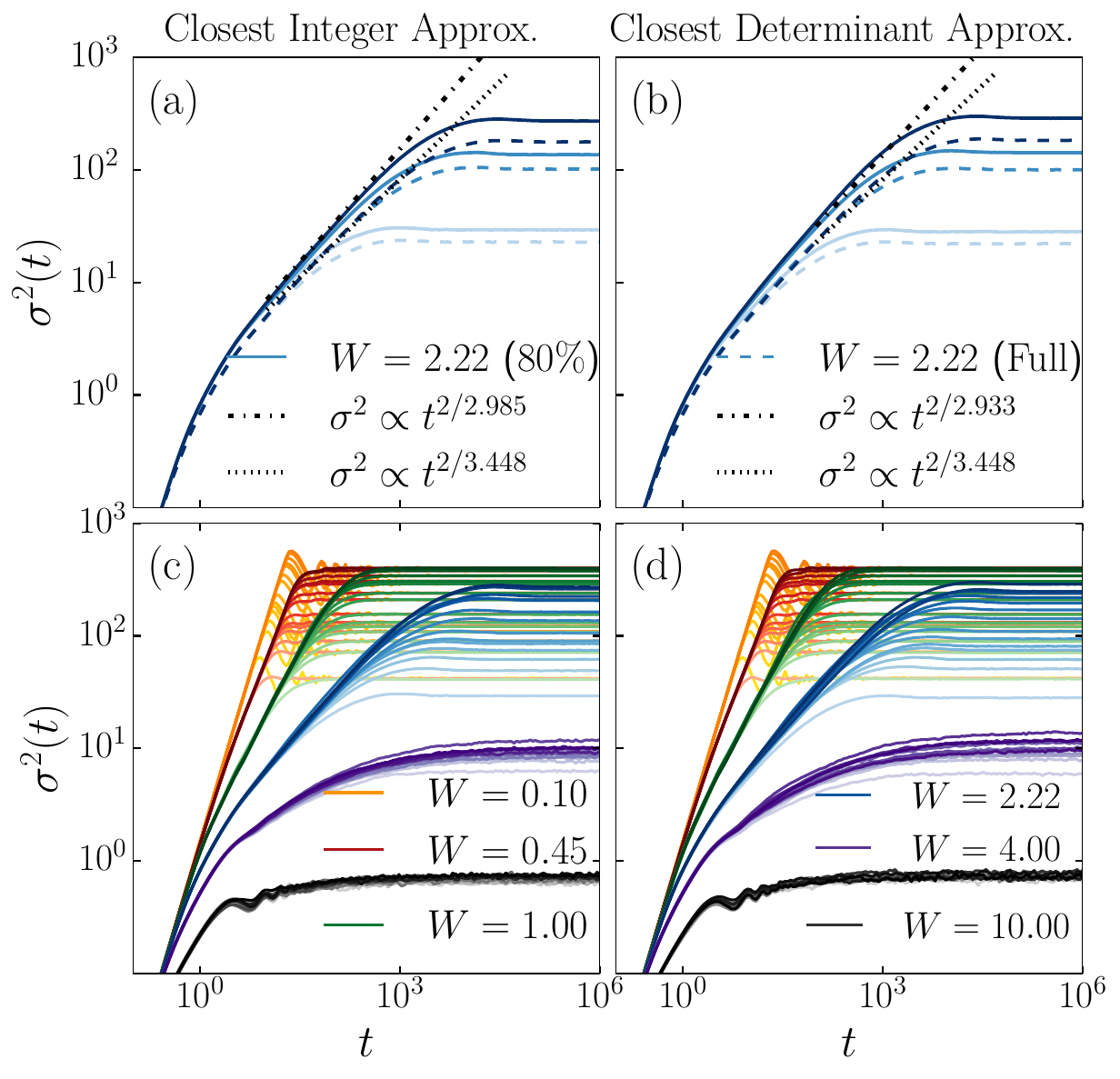}
  \caption{The mean square displacement $\sigma^2_{}(t)$ in the 3D Aubry-André model with $\theta=\pi/7$ and $\beta=(\sqrt{5}-1)/2$ in (a), (c) the closest integer approximation and (b), (d) the closest determinant approximation of the $B$-matrix. We consider $10$ realizations and for each realization we consider $100$ different initial positions. (a)-(b) Comparison of the average over initial stated with ($80\%$) and without (Full) the constraint of Eq.$~$\eqref{eqn:projection2} for 
  selected system sizes $L=13,25,40$. (c)-(d) Comparison of selected representative potential strength $W$ and for all system sizes considered here, i.e., 
$L\in\{10,11,13,17,19,21,22,23,25,29,31,34,35,37,39,40\}$, from light to dark gradient of each color.
  }
  \label{fig:msd}
\end{figure}

In the rest of this work, we discuss only the selective averaging.
In Figs.$~$\ref{fig:msd}(c) and$~$\ref{fig:msd}(d), we show $\sigma^2_{}(t)$ for several representative potential strengths covering the real-space delocalized phase, $W=0.1,0.45,1$, the representative value of the critical point $W=2.22$, and the representative values of the real-space localized phase $W=4,10$. With increasing $W$ within the delocalized phase, the hydrodynamic slope decreases. In the localized phase, i.e., $W=4$ and $10$, the mean square displacement saturates to system-independent curves.

The hydrodynamic mid-time slope of the mean square displacement
can be quantified by the dynamical exponent $z$$~$\cite{hopjan2025critical} defined via the power-law scaling,
\begin{equation}
    \sigma^2(t) \propto t^{2/z}.
\end{equation}
For ballistic transport, $z=1$, which is approximately observed for $W=0.1$, see the yellow curves in Figs.$~$\ref{fig:msd}(c) and$~$\ref{fig:msd}(d). In the case of diffusion, $z=2$, which is approximately observed for $W=1$, see the green curves in Figs.$~$\ref{fig:msd}(c) and$~$\ref{fig:msd}(d). Furthermore, for the super-diffusive case, $2>z>1$, and for the sub-diffusive case, $z>2$. Finally, in the localized case, $z\to \infty$, which is observed for $W>4$, see the purple and black curves in Fig.$~$\ref{fig:msd}(c) and$~$\ref{fig:msd}(d). The careful extraction of the dynamical exponent is provided in Sec.$~$\ref{3D case time dynamical exponent}.

\subsection{Scaling of steady-state values and their fluctuations}
\label{3D case time steady state}

In order to determine from the dynamics whether the state is delocalized or localized, it is instructive to study the system size dependence of the saturation value of $\sigma(t)$, i.e., the value of $\sigma(t)$ after the Thouless time
\begin{equation}
 \sigma^2(t\rightarrow \infty) = \sigma^2_{\text{ss}}.
\label{eqn:msd}
\end{equation}
In our calculations, late times ranges from $t=10^5$ to $10^6$, such that $\sigma^2(t)$ reaches its saturation value for all potential strengths and all system sizes considered here.

\begin{figure}[t]
    \includegraphics[width=0.45\textwidth]{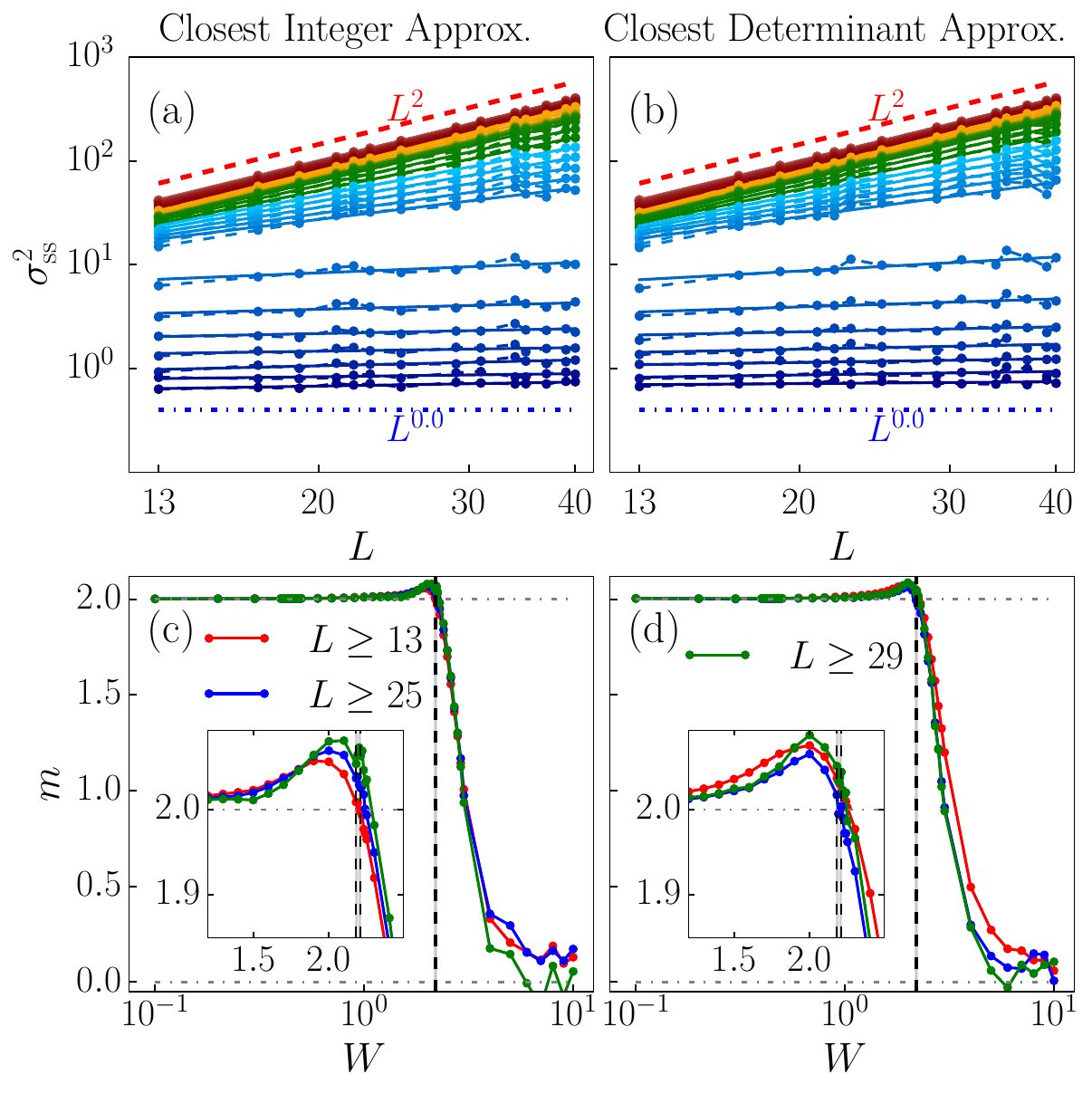}
  \caption{(a) and (b) $\sigma^2_{\text{ss}}$ which is the average of $\sigma^2(t)$ in times between $10^5$ and $10^6$, as a function of the system size $L$ and various $W$ values (red to blue color) in the 3D Aubry-André model for $\theta=\pi/7$ and $\beta=(\sqrt{5}-1)/2$ in (a) the closest integer approximation and (b) the closest determinant approximation of the $B$-matrix.  We consider $10$ realizations and for each realization we consider $100$ different initial positions. (c) and (d) Slopes $m$ extracted from the fits in panels (a) and (b), respectively, as functions of $W$. We extract the slope $m$ considering three sets of linear system sizes: (i) $L\geq13$ (red), (ii) $L\geq 25$ (blue), (iii) $L\geq29$ (green). Insets of (c) and (d): details around the transition point $W_c$, where $m\approx2$ values start to drop to $0$.  The vertical dashed lines denote the transition point extracted in Sec.$~$\ref{3D case universal}, $W_c^{}\approx2.18-2.21$. }
  \label{fig:sigma_ss_avg}
\end{figure}

In Figs.$~$\ref{fig:sigma_ss_avg}(a) and$~$\ref{fig:sigma_ss_avg}(b), we plot $\sigma^2_{\text{ss}}$ as a function of $L$ for several system sizes in the range $L=13-40$, and extract the slope $m$ from fitting to the ansatz 
\begin{equation}
\sigma^2_{\text{ss,fit}} \propto L^m.
\label{eqn:msd}
\end{equation}
For the fitting, we consider three sets of system sizes: the sets with all system sizes satisfying $L\geq 13$, $L\geq 25$, and $L\geq 29$, respectively. For each set separately, we extract the slope.

In Figs.$~$\ref{fig:sigma_ss_avg}(c) and$~$\ref{fig:sigma_ss_avg}(d), we plot the resulting $m$ as a function of $W$. Remarkably, we observe that $m\approx2$ for $W\leq W_c$, and just at the estimated transition point, $m$ starts to decrease rapidly to zero, see the insets of Figs.$~$\ref{fig:sigma_ss_avg}(c) and$~$\ref{fig:sigma_ss_avg}(d) for details. This suggests that the selected initial states indeed spread throughout the whole system, reflecting the delocalization of the eigenstates to which the selected initial states mostly project.
For $W\geq 4$, there are clear signatures of localization as $m\approx0$. The behavior for $W_c<W<4$ can be considered a finite-size crossover regime. In the thermodynamic limit, we expect a step-like structure with $m=2$ for $W \leq W_c$, and $m=0$ for $W > W_c$. Finally, in Appendix$~$\ref{3D case transport},  we study the system-size fluctuations of $\sigma^2_{\text{ss}}$ above $\sigma^2_{\text{ss,\ fit}}$ and show that they are similar to the fluctuations in eigenstate properties.

\subsection{Extraction of dynamical exponent}
 \label{3D case time dynamical exponent}

We now turn to the last step of the transport characterization, the extraction of the dynamical exponent.
To do so, we study the logarithmic derivative of the mean square displacement, which can be interpreted as the time-dependent dynamical exponent$~$\cite{Sierant2020},
\begin{equation}
    \frac{1}{z(t)} = \frac{1}{2}\frac{d\log\sigma^2(t)}{d\log t}.
\end{equation}
To extract $1/z$ in the mid-time hydrodynamical regime, for each $W$, we identify a plateau of finite length, starting from some initial time $t_{\text{min}}$ to some final time $t_{\text{max}}$ and calculate the average value of $1/z$ in this time interval. 
We note that there is no unique way to choose the width of the time interval of the plateau, i.e., $t_{\text{min}}$ and $t_{\text{max}}$, and it is done empirically here.
 
\begin{figure}[t]
  \includegraphics[width=0.45\textwidth]{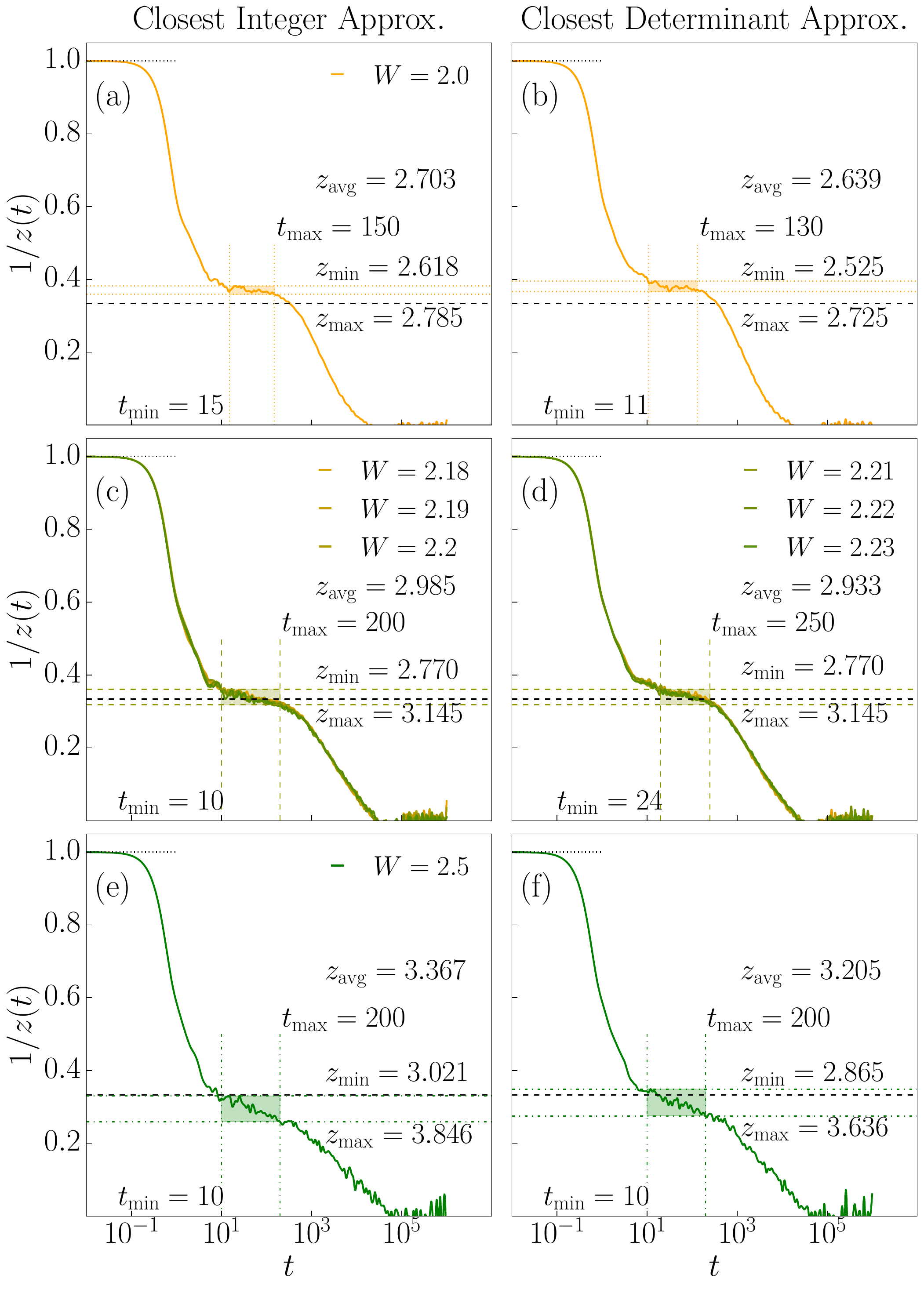}
  \caption{The time dependent dynamical exponent $1/z(t)$ as a function of time $t$ for selected $W$ and $L=40$ in
  the 3D Aubry-André model for $\theta=\pi/7$ and $\beta=(\sqrt{5}-1)/2$ in the closest integer approximation of the $B$-matrix, panels (a), (c) and (e), and in the closest determinant approximation of $B$-matrix, panels (b), (d) and (f). The colored vertical lines denote the beginning and the end of the hydrodynamical time interval $t_{\text{min}}$ and $t_{\text{max}}$. The colored horizontal lines denote the maximal $(1/z)_{\text{max}}$ and the minimal $(1/z)_{\text{min}}$ values of $1/z(t)$ in the hydrodynamical time interval. The black dashed horizontal lines denote the theoretical prediction $z=3$ for the critical point.}
  \label{fig:alpha}
\end{figure}

In Fig.$~$\ref{fig:alpha}, we illustrate this procedure for several representative values of $W$, i.e., $W=2$ representing the delocalized phase, $W=2.18-2.23$ representing the vicinity of the critical point, and $W=2.5$ representing the localized phase. To account for uncertainties, we consider a finite range of $1/z$, i.e., the maximal $(1/z)_{\text{max}}$ and the minimal $(1/z)_{\text{min}}$ values. The most representative value is an average over all $1/z$ in the interval $[t_{\text{min}},t_{\text{max}}]$, denoted as $(1/z)_{\text{avg}}$.

\begin{figure}[t]
  \includegraphics[width=0.47\textwidth]{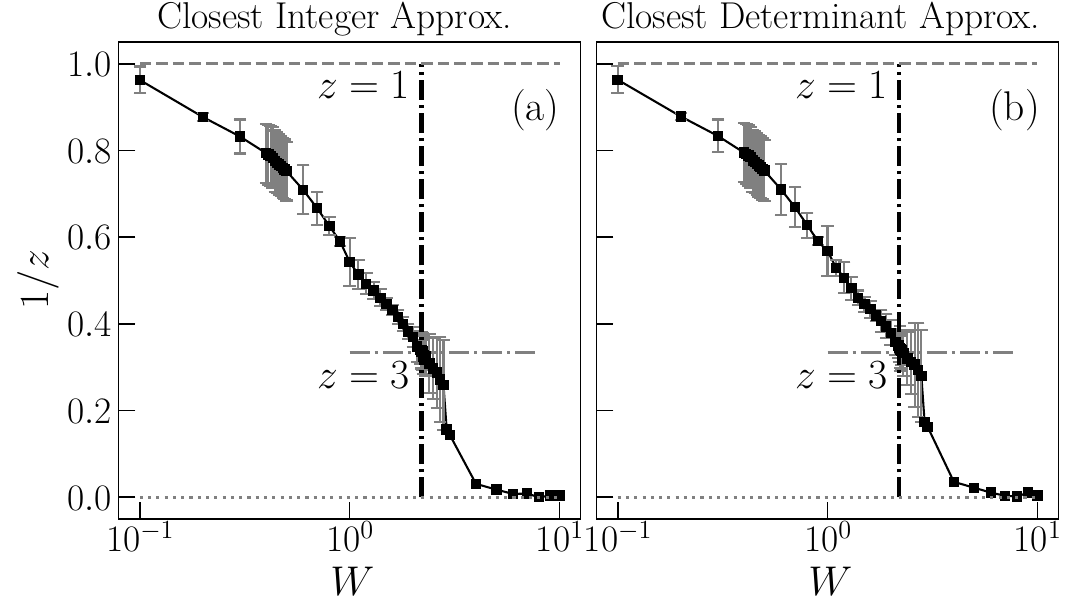}
  \caption{The extracted dynamical exponent $1/z$ as a function of $W$ in the 3D Aubry-André model for $\theta=\pi/7$ and $\beta=(\sqrt{5}-1)/2$ in (a) the closest integer approximation of $B$-matrix and (b) the closest determinant approximation of $B$-matrix.}
  \label{fig:alpha_m}
\end{figure}

The extracted values of $(1/z)_{\text{avg}}\equiv 1/z$ as a function of $W$ are shown in Fig.$~$\ref{fig:alpha_m}. We estimate the error bars of $1/z$ as the difference between $1/z_{\text{min}}$ and $1/z_{\text{max}}$. We observe that $1/z$ is continuously decreasing from the ballistic value, see the horizontal gray dashed lines in Fig.$~$~\ref{fig:alpha_m}, to a value $1/z\approx 1/3$ at $W_c$, see the gray horizontal dash-dotted lines in Fig.$~$~\ref{fig:alpha_m}. For $W=2.5>W_c$, the identification of the plateau in $1/z(t)$ becomes harder; see Figs.$~$\ref{fig:alpha}(e) and$~$\ref{fig:alpha}(f). Thus, the extracted values of $1/z$ suffer from large uncertainties, see Fig.$~$\ref{fig:alpha_m}. For $W>4$, where there is a lack of the hydrodynamical regime, $\sigma^2_{}(t)$ saturates to system-independent profile, which is equivalent to $1/z=0$, see Fig.$~$\ref{fig:alpha_m}. Interestingly, we do not observe any clear indication of a diffusive phase between $1/W_c$ and $W_c$, as suggested in Ref.$~$\cite{devakul2017anderson}, which calls for further study.

The values of the dynamical exponent $z$ are generally not possible to predict. The only exception is the critical point, where the dynamical properties, such as the survival probability$~$\cite{hopjan2023} and mean square displacement$~$\cite{hopjan2025critical}, can be linked to the spectral properties$~$\cite{hopjan2023,hopjan2025critical}. Then the dynamical exponent has a particularly simple prediction$~$,\cite{hopjan2025critical}
\begin{equation}
    z=\frac{d_l}{d_s}
\end{equation}
where $d_l$ is the lattice dimension and $d_s$ is the spectral fractal dimension. We note that the derivation of the critical dynamical exponent, as presented in Ref.$~$\cite{hopjan2025critical}, relies on the assumption that $\sigma^2_{\text{ss}} \propto L^2$.  This assumption was checked in Sec.$~$\ref{3D case time steady state}. 

Thus, with $d_l=3$ and $d_s=1$ (see Appendix$~$\ref{3D case d_s} for evidence), the critical dynamical exponent in the present model is expected to be subdiffusive with a value $z=3$. In Fig.$~$\ref{fig:alpha_m}, we indeed find a very good agreement between this expectation and the actual result. The estimated values of $z$ are reported in Figs.$~$\ref{fig:alpha}(c) and$~$\ref{fig:alpha}(d). For $W\approx W_c$, we get $z \approx 2.985$ for the closest integer approximation, and $z\approx 2.933$ from the closest determinant approximation. Along with the test of the universal relation between the eigenstate and spectra, discussed in Sec.~\ref{3D case universal}, this is another confirmation of the consistency of our transition point estimate.

\section{Conclusions} \label{conclusions}

We have carried out a comprehensive numerical study of the 3D Aubry-André model, focusing on its static eigenspectral and eigenstate properties, as well as its dynamical properties across the localization transitions. By combining spectral diagnostics, wave-function measures, and dynamical observables, we established a consistent picture of the phase diagram, see Fig.$~$\ref{fig:intro}, and the associated critical behavior.

The duality of the model gives rise to two distinct transition points related by the transformation $W \leftrightarrow 1/W$. The transition at $W_c^{}\approx2.18 - 2.21$ corresponds to a localization transition in real space, while its dual counterpart at $1/W_c^{}$ corresponds to localization in the dual momentum space. Between these two critical points lies a truly delocalized ergodic phase, whereas outside this interval, the eigenstates are localized in one of the spaces. This demonstrates that the 3D Aubry-André model realizes localization transitions at $1/W_c$ that have no direct analog in the paradigmatic 3D Anderson model, where only real-space localization at $W_c$ occurs$~$\cite{Lydzba26}. That being said, the value $\nu \approx 1.502-1.507$ of the correlation length exponent at $W_c$ is very close to the value $\nu \approx 1.57$ in the 3D Anderson model$~$\cite{Slevin_2014}, hinting at certain similarities between the two transitions.

We note that the emergence of the truly delocalized phase between the two phases with Poissonian statistics, when compared to the 1D Aubry-André model, can be attributed to the dimensionality of the 3D Aubry-André model. This resembles the situation in the Anderson model. The 1D Anderson model is localized, and delocalization can be caused by increased dimensionality. However, there exists another possibility to induce the delocalization-localization transition, even in 1D,  by adding a long-range coupling between lattice sites, as in the Power-Law Random Banded model$~$\cite{Mirlin96, Mirlin00,Evers00}. Applying this analogy to the 1D Aubry-André model, one may expect an emergent phase between two phases with Poissonian statistics when considering a 1D self-dual model with long-range coupling. Indeed, for a class of self-dual models introduced in Ref.$~$\cite{Nosov19}, there exists a phase exhibiting short-range Wigner-Dyson level statistics.

A major challenge in studying finite-size scaling in the 3D Aubry-André model is the presence of strong finite-size fluctuations originating from the imposition of the commensurate approximation. We show that these fluctuations can be substantially reduced by averaging over the central $50\%$ of the spectrum, allowing for reliable finite-size scaling analyzes. Independent analyzes based on the gap ratio, inverse participation ratio, spectral form factor, participation entropy, and spectral compressibility consistently locate the transition near $W_c^{}\approx2.18 - 2.21$. The spectral statistics evolve from GUE in the ergodic regime to Poisson statistics in the localized phases, while intermediate statistics emerge at criticality.

\begin{figure}[t!]
    \includegraphics[width=0.45\textwidth]{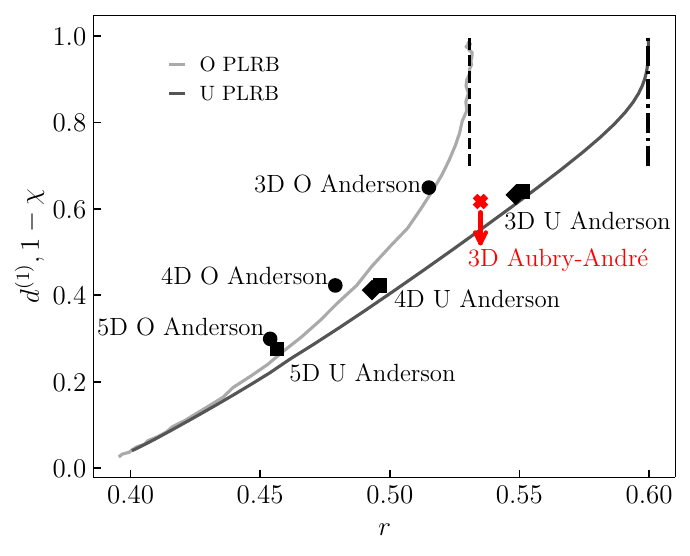}
  \caption{The critical 3D Aubry-André model at the landscape of critical models of Ref.$~$\cite{jiricek2025universal}, consisting of the orthogonal (O) and unitary (U) 3D, 4D and 5D Anderson models (black filled symbols) and the orthogonal and unitary Power-Law Random Banded (PLRB) models (light and dark gray solid lines, respectively). The dashed and dash-dotted lines indicate the limiting values of the gap ratio$~$\cite{atas_bogomolny_13}, $r\approx 0.5307$ and $r\approx 0.5996$, for the GOE and GUE, respectively. The black filled symbols and the gray solid lines are results from Ref.$~$\cite{jiricek2025universal}, while the red cross is the new result of this work. }
  \label{fig:landscape_3D}
\end{figure}

Beyond identifying the transition points, we demonstrated that both critical points exhibit the universal characteristics expected for localization transitions between quantum ergodicity and localization. In particular, the spectral compressibility and the fractal dimension of the critical eigenstates satisfy the recently proposed universal relation$~$\cite{jiricek2025universal}. In Fig.$~$\ref{fig:landscape_3D}, we compare the critical static properties of the 3D Aubry-André model to the critical static properties of the models discussed in Ref.$~$\cite{jiricek2025universal}. We observe that the 3D Aubry-André model in our implementation follows, with good accuracy, the analytical curve $r(\chi)$ obtained in Ref.$~$\cite{jiricek2025universal} from the unitary Power-Law Random Banded (PLRB) model$~$\cite{Mirlin96}. Thus, the 3D Aubry-André model belongs to the class of critical unitary models, along with the unitary Anderson models.

Going beyond the static properties, the wave-packet dynamics at the transition $W_c$ are governed by the critical dynamical exponent$~$\cite{hopjan2025critical} with a sub-diffusive value $z=3$, providing complementary dynamical evidence for criticality. This is another similarity between the critical properties of $W_c$ in the 3D Aubry-André model and the critical properties of the critical point in the 3D Anderson model, where the critical dynamical exponent is also sub-diffusive with the value of $z=3$. While we have explicitly demonstrated the critical properties only for $W_c$, the critical properties of the second transition point at $1/W_c$ result implicitly from the model symmetry, which was thoroughly discussed in this work.

Our work establishes the 3D Aubry-André model as a unique platform for studying critical phenomena not only in real space but also in dual momentum space, and it can serve as a starting point in the investigation of the ergodic properties of higher-dimensional quasiperiodic systems.

\section{Acknowlegment}
We acknowledge support from the Slovenian Research and Innovation Agency (ARIS), Research core funding Grants No.~P1-0044, N1-0273 and J1-50005. M. H. acknowledges support from the Polish National Agency for Academic Exchange (NAWA)’s Ulam Programme (project BNI/ULM/2024/1/00124). We gratefully acknowledge the High Performance Computing Research Infrastructure Eastern Region (HCP RIVR) consortium$~$\cite{vega1} and European High Performance Computing Joint Undertaking (EuroHPC JU)$~$\cite{vega2}  for funding this research by providing computing resources of the HPC system Vega at the Institute of Information sciences$~$\cite{vega3}.

\twocolumngrid
\appendix
\label{app}

\section{Full vs diagonal averaging of commensurate approximation}\label{fulldiag}

\begin{figure}[t!]
    \includegraphics[width=0.45\textwidth]{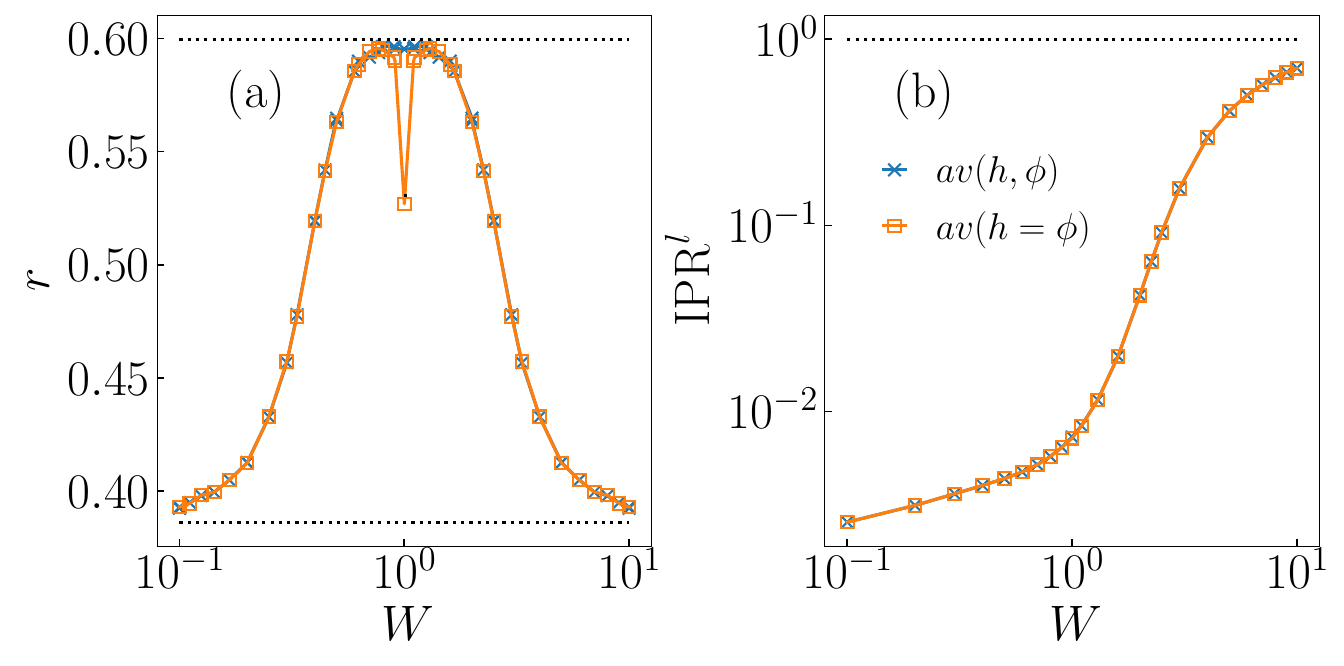}
  \caption{Averaged (a) gap ratio $r$  and (b) real-space inverse participation ratio ${\rm IPR}^l$, as a function of system size $W$ for $L=8$ in the closest determinant approximations of the 3D Aubry-André model. We show comparison of the averages over $1000$ randomly selected pairs $(\vec{h},\vec{\phi})$ with the averages over $1000$ randomly selected diagonal elements $\vec{h}=\vec{\phi}$ of the Brillouin zone. For each realization, we average over bulk $50\%$ states.}
  \label{fig:averaging_3D}
\end{figure}

In this section, we study the difference between the averaging over randomly selected pairs of $(\vec{h},\vec{\phi})$ from the Brillouin zone and the averaging over randomly selected phases $\vec{\phi}$ with the Bloch number, having the same value $\vec{h}=\vec{\phi}$; see Sec.$~$\ref{commensurate approx} and Sec.$~$\ref{diag commensurate approx}, respectively. In Fig.$~$\ref{fig:averaging_3D} we plot the average gap ratio and the average IPR, see Sec.$~$\ref{3D case scaling} for the definitions, as a function of $W$ for rather small linear size $L=8$.

For the gap ratio, we observe almost no difference between the full and the diagonal averaging. The only exception is the vicinity of the selfdual point $W=1$, which lies in the ergodic phase. On the one hand, the average gap ratio from the full averaging is $r\approx 0.5996$. On the other hand, the diagonal averaging gives $r\approx 0.5307$. The former and latter values are theoretical predictions for the systems of the Gaussian orthogonal ensemble (GOE) and Gaussian unitary ensemble (GUE), respectively, i.e., for ergodic systems with level repulsion. The source of this difference is hidden in the difference between the Hamiltonian in Eq.$~$\eqref{eqn:H_3DAA_real_comm} and in Eq.$~$\eqref{eqn:H_3DAA_real_2}. 
The latter fulfills the time-reversal symmetry at $W=1$ and therefore belongs to the GOE universality. Away from the vicinity of the selfdual point, the results are practically indistinguishable. 
This is further supported by the comparison of the IPRs in Fig.$~$\ref{fig:averaging_3D}(b). They are practically indistinguishable over the entire range of the parameters. Thus, as in the region of the eigenstate transition, which is the main focus here, the two averages give quantitatively similar values for both the spectral and wave-function measures; we use the diagonal averaging.

\section{System-size fluctuations in spectral measures}
 \label{3D case fluctspec}

\begin{figure}[t]
  \includegraphics[width=0.45\textwidth]{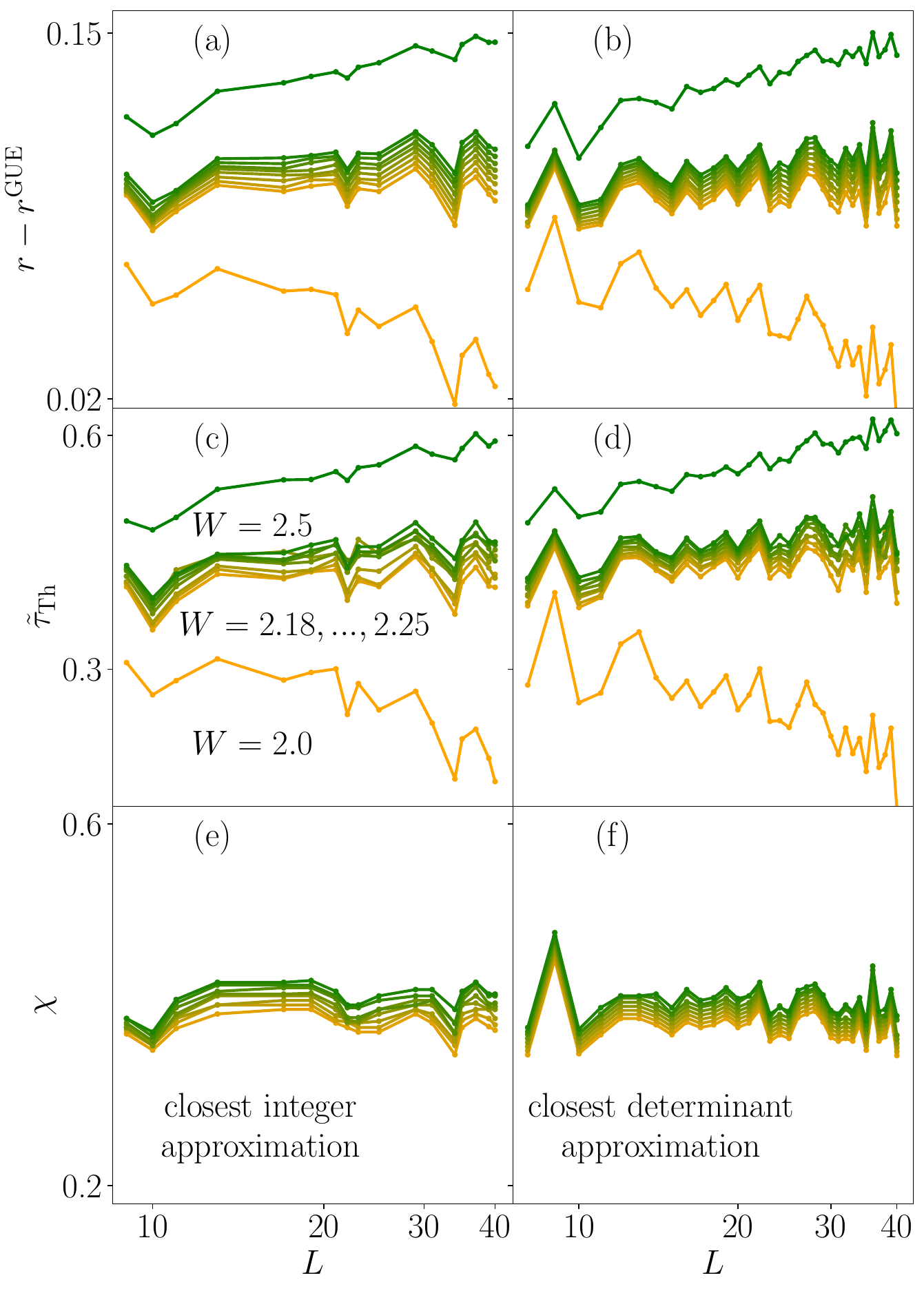}
  \caption{The spectral properties of the 3D Aubry-André model for $\theta=\pi/7$ and $\beta=(\sqrt{5}-1)/2$ in (a), (c), (e) the closest integer approximation of the $B$-matrix and (b), (d) and (f) the closest determinant approximation of the $B$-matrix. The comparison of the finite-size dependence of (a)-(b) $r_{\rm GUE}-r$, (c)-(d) the Thouless time $\tilde{\tau}_{{\rm Th}}$, and (e)-(f) the compressibility $\chi$. The gap ratio $r$ is the averaged over $50\%$ eigenstates around the mid spectra, and the Thouless time and the compresibility are extracted from the normalized SFF with the smooth box function with $\eta_{\rm aux}=0.1$. We average over $100$ Hamiltonian realizations.
  }
  \label{fig:r_tau_scaling}
\end{figure}

In this section, we focus on the system size dependence of the spectral indicators, i.e., the average gap ratio, the normalized Thouless time, and the spectral compressibility, for several values of $W$ close to the transition. We consider both approximations from Sec.$~$\ref{choice dual approx}, i.e., the closest integer and the closest determinant approximation, and we study whether our conclusions depend on the particular approximation. Both approximations are shown in parallel in  Fig.$~$\ref{fig:r_tau_scaling}. Figures$~$\ref{fig:r_tau_scaling}(a), $~$\ref{fig:r_tau_scaling}(c), and $~$\ref{fig:r_tau_scaling}(e) correspond to the closest integer approximation, while Figs.$~$\ref{fig:r_tau_scaling}(b), $~$\ref{fig:r_tau_scaling}(d), and $~$\ref{fig:r_tau_scaling}(f) correspond to the closest determinant approximation. In Figs.$~$\ref{fig:r_tau_scaling}(a) and$~$\ref{fig:r_tau_scaling}(b), we show the system size dependence of the averaged gap ratio, more specifically its distance to the GUE value, $r -r^{{\rm GUE}}$. In Figs.$~$\ref{fig:r_tau_scaling}(c) and$~$\ref{fig:r_tau_scaling}(d), we further show  the system size dependence of the normalized Thouless time $\tilde{\tau}_{{\rm th}}$. Finally, Figs.$~$\ref{fig:r_tau_scaling}(e) and$~$\ref{fig:r_tau_scaling}(f) contain values of the spectral compressibility $\chi$. 

We observe that, close to the transition $W\in[2.18,2.25]$, all the spectral indicators exhibit saturation as a main trend. However, on top of the main trend, we observe additional system size fluctuations. On the one hand, different system size fluctuations are present in the closest integer approximation compared to the closest determinant approximation. This is not unexpected, as the approximations for the same system size differ by construction. On the other hand, the main trends seem to be very similar even for different commensurate approximations; compare the main trends in the left and right columns of Fig.$~$\ref{fig:r_tau_scaling}. Remarkably, the fluctuations of all spectral indicators possess similar profiles within each of the commensurate approximations, i.e., within each column in Fig.$~$\ref{fig:r_tau_scaling} we observe similar profiles. This suggests that the presence of the  system-size fluctuations is systematic.

Finally, in Figs.$~$\ref{fig:fluct}(a) and$~$\ref{fig:fluct}(b), we also plot the differences between the values of the spectral compressibility and its mean value, i.e., $\chi-\chi^{{\rm fit}}$, as a function of system sizes. The latter was obtained from the fitting of $\chi$ in Sec.$~$\ref{3D case universal}. We observe that the fluctuations above the main trend are very similar for all $W$ close to the estimated transition point.

\section{System-size fluctuations in wave-function measures}
 \label{3D case fluctwave}

In this section, we turn to the wave-function properties of the eigenstates. We ask whether they exhibit systematic system-size fluctuations that are similar to those observed in the spectral measures. As motivated in Sec.$~$\ref{3D case universal}, we characterize the wave-function properties with the participation entropies.

\begin{figure}[t]
\includegraphics[width=0.45\textwidth]{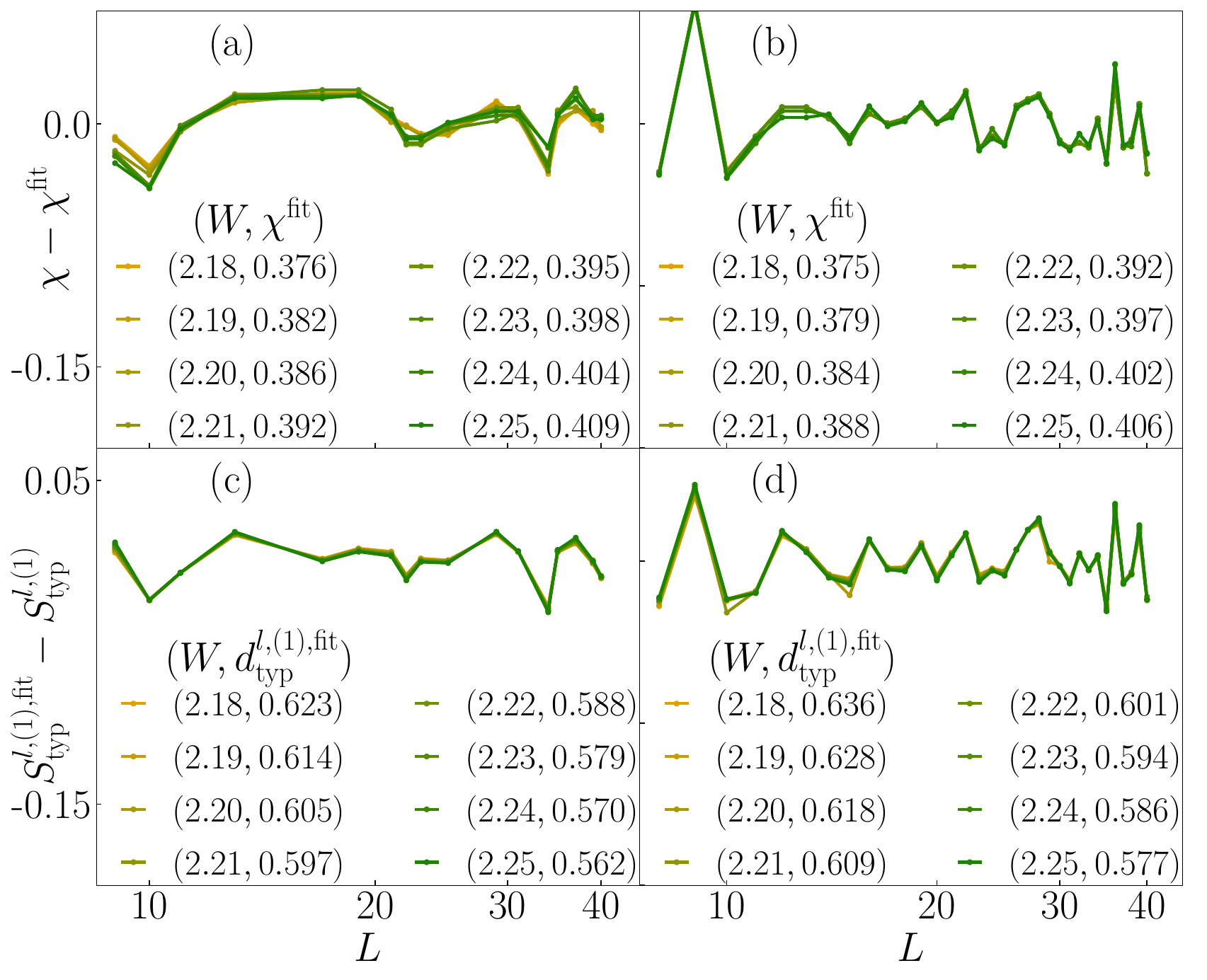}
  \caption{Fluctuations of the spectral and wave-function measures above the main trends in the 3D Aubry-André model for $\theta=\pi/7$ and $\beta=(\sqrt{5}-1)/2$ in (a), (c) the closest integer approximation of the $B$-matrix and, (b), (d) in the closest determinant approximation of the $B$-matrix. Panels (a)-(b) show the fluctuations of the spectral compresibility above its mean $\chi-\chi^{{\rm fit}}$, and panels (c)-(d) show the fluctuations of the participation entropy $S_{\rm typ}^{l,(1)}$ above linear fit $S_{\rm typ}^{l,(1),{\rm fit}}$.
  }
  \label{fig:fluct}
\end{figure}

In order to get comprehensive information of the wave-function, we study not only the second moment of the wave-function amplitudes, i.e., the IPR, but also $q$-moments of the wave-function amplitudes.
We consider the IPR in the lattice basis $b=l$, see Eq.$~$\eqref{eq:IPR_def}, and its generalization to an arbitrary index $q$$~$ defined as$~$\cite{suntajs2024} 
\begin{equation}
    {\rm IPR}_{\mu,H}^{l,(q)} = \sum_{l=1}^{L^{3}} |c_{\mu,H}^l|^{2q},
\end{equation}
which for $q=2$ reduces to the IPR in Eq.$~$\eqref{eq:IPR_def}. Then, the participation entropy is defined as
\begin{equation}
    S_{\mu,H}^{l,(q)} = \frac{1}{1-q}\ln \big({\rm IPR}_{\mu,H}^{l,(q)}\big),
    \label{eqn:Sq}
\end{equation}
and its $q=1$ limit is known as the von Neumann participation entropy,
\begin{equation}
    S_{\mu,H}^{l,(1)} =  -\sum_{l=1}^{V} |c_{\mu,H}^l|^{2}\ln \big(|c_{\mu,H}^l|^{2}\big).
\end{equation}
The advantage of the participation entropy over the IPR is the possibility of measuring the fractal dimension for $q=1$ from its slope$~$\cite{suntajs2024}.

Thus, we focus on the participation entropies for the rest of the work. Similarly to the IPR, we study the averaged participation entropy defined as
\begin{equation}
    S^{l,(q)}_{{\rm typ}}
    = \Bigg\langle  \frac{1}{N_{C}} \sum_{\mu=N_{{\rm i}}}^{N_{{\rm i}} + N_{C}-1} S_{\mu,H}^{l,(q)}\Bigg\rangle_H.
    \label{eqn:Sq_typ}
\end{equation} 
Here, as in the previous sections, we consider $50\%$ eigenstates around the mid spectra. We note that the average is performed outside of the logarithm. For $q=2$, the participation entropy can be interpreted as the logarithm of the typically averaged IPR. We thus refer to the participation entropy as the typical participation entropy. 

\begin{figure}[t]
\includegraphics[width=0.45\textwidth]{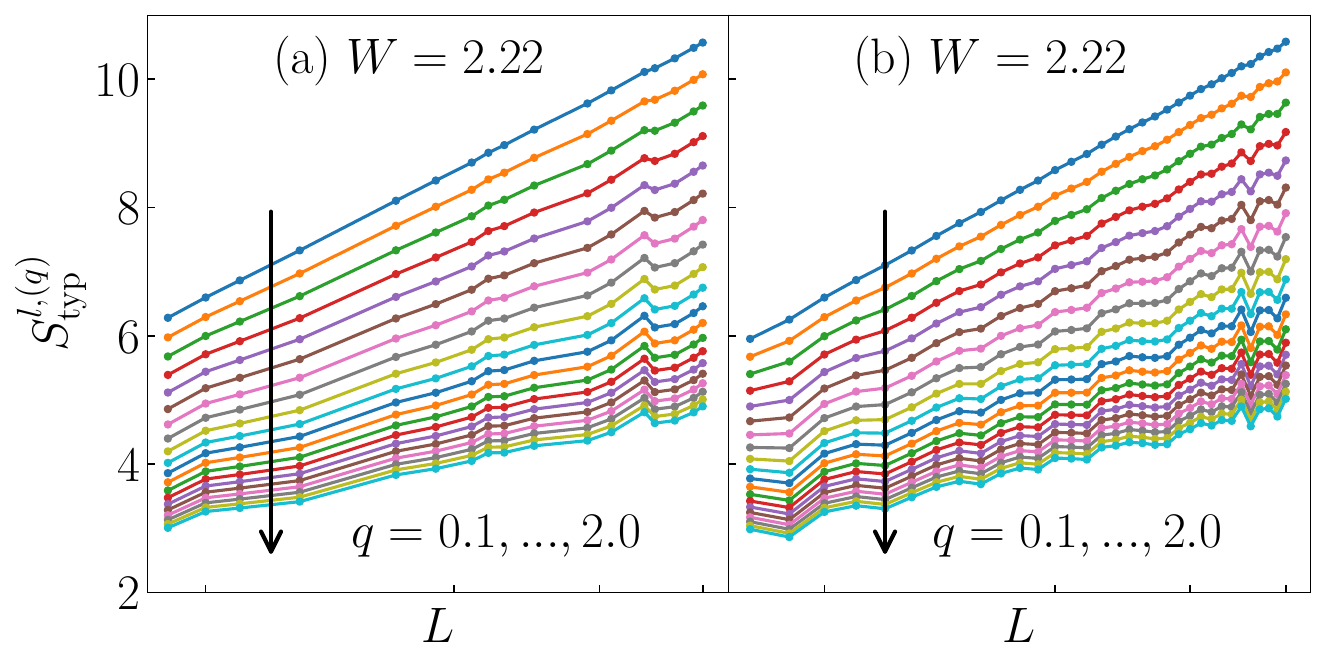}
  \caption{The wave-function properties of the 3D Aubry-André model for $\theta=\pi/7$ and $\beta=(\sqrt{5}-1)/2$ in (a) the closest integer approximation of the $B$-matrix and, (b) the closest determinant approximation of $B$-matrix. Examples of the system-size dependence of the participation entropy $S_{\rm typ}^{l,(q)}$ for different moments $q$ for selected value of the potential strength $W$. We average over $50\%$ eigenstates around the mid spectra and $100$ Hamiltonian realizations.
  }
  \label{fig:Sq}
\end{figure}

In Figs.$~$\ref{fig:Sq}(a) and$~$\ref{fig:Sq}(b), we plot the typical participation entropy $S^{l,(q)}_{{\rm typ}}$ as a function of system size $L$ for different $q$ values ranging from $0.1$ to $2.0$ at the representative value $W=2.22$, which is close to the estimated transition. We observe that the typical participation entropy is generally an increasing function, with a main trend that can be described as linear growth with the logarithm of the volume, i.e., $\ln (V) =\ln (L^3)$. Similar to the spectral measures, we observe additional fluctuations on top of the main trend.

To characterize the fluctuations, in Figs.$~$\ref{fig:fluct}(c) and$~$\ref{fig:fluct}(d), we plot the difference between the participation entropy $S^{l,(1)}_{{\rm typ}}$ and its fit $S^{l,(1),{\rm fit}}_{{\rm typ}}$. The latter was obtained with the fit to the function from Eq.$~$\eqref{eqn:Sq_fit} in Sec.$~$\ref{3D case universal}. We first observe that the fluctuations above the main trend are very similar for all $W$. More importantly, the profile of the fluctuations in the wave-function measures is qualitatively similar to the profile of the fluctuations in the spectral measures, compare Fig.$~$\ref{fig:fluct}(a) to Fig.$~$\ref{fig:fluct}(c) and Fig.$~$\ref{fig:fluct}(b) to Fig.$~$\ref{fig:fluct}(d). For other $q$-values the picture is similar (not shown). We can thus conclude that the system-size fluctuations are systematically present not only in the spectral measures, but also in the eigenstate measures, such as the participation entropies.

\section{Avoiding effects of spectral edges in transport}
\label{3D case time initial state}

As discussed in Sec.$~$\ref{3D case edges}, the transition from ergodicity to localization can be detected from $50\%$ of energy levels around the middle of the spectrum. However, not all initial states reflect the physics of the midspectra in their dynamics. Only the initial states that have a large projection to the midspectra do so. 

\begin{figure}[t!]
\includegraphics[width=0.45\textwidth]{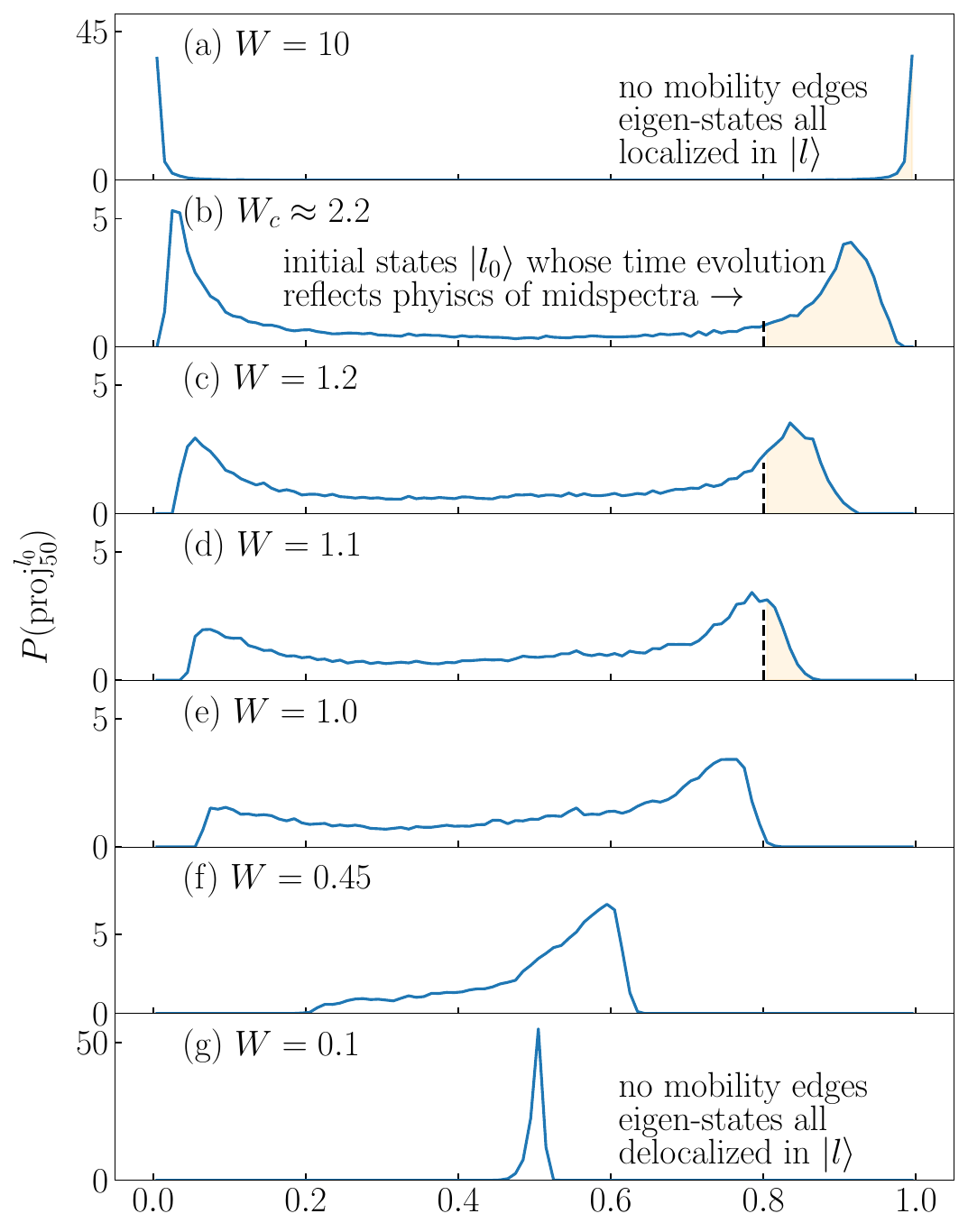}
\caption{Probability $P({\rm proj}_{50}^{l_0})$ of the net projection ${\rm proj}_{50}^{l_0}$ of the initially localized states $|l_0\rangle$ to mid 50$\%$ of the eigenstate for several values of $W$ representing phases: (a) localized, (b) critical and (c)-(g) delocalized in real space. The yellow area denotes states which probes transport properties of the mid-spectra eigenstates.} 
\label{fig:initial_proj}
\end{figure}

To separate the suitable initial states, it is instructive to study a net projection of the initial states $|l_0\rangle$ to the midspectrum, i.e.,
\begin{align}
{\rm proj}_{50}^{l_0}=\sum_{\mu=V/4}^{3V/4} |\langle \mu | l_0\rangle|^{2},
\label{eqn:projection}
\end{align}
and their histograms. An example of the histogram of ${\rm proj}_{50}^{l_0}$, averaged over 100 Hamiltonian realizations, is shown in Fig.$~$\ref{fig:initial_proj} for several values of $W$.

The structure of such a histogram can be understood when approaching it from the localized limit. At $W=10$, the histogram is double peaked, see Fig.$~$\ref{fig:initial_proj}(a), meaning that $|l_0\rangle$ projects either to the mid-states or edges. Here, all eigenstates are localized, and each of the initial states $|l_0\rangle$ is associated with the localization center of some eigenstate $|l_0\rangle \leftrightarrow |\mu\rangle$. Thus the state $|l_0\rangle$ mostly projects to only one eigenstate, and such an eigenstate can be either in the bulk of the spectrum, which corresponds to the upper peak in Fig.$~$\ref{fig:initial_proj}(a), or at the edges, which corresponds to the lower peak in Fig.$~$\ref{fig:initial_proj}(a).

At the transition point, $W_c\approx2.2$, the mid-spectrum eigenstates become critical, while the edge eigenstates remain localized. The mobility edge develops. Consequently, the initial states $|l_0\rangle$, which mostly project into the mid-spectrum eigenstates, become mobile in the lattice. We are mainly interested in these initial states as they exhibit signs of critical transport. We set quite a strict criterion for the initial state $|l_0\rangle$ to be considered critical, i.e.,
\begin{align}
{\rm proj}_{50}^{l_0}>0.8,
\label{eqn:projection2}
\end{align}
see the yellow upper peak in Fig.$~$\ref{fig:initial_proj}(b). 

Interestingly, we observe that the double-peak structure in $P({\rm proj}_{50}^{l_0})$ survives down to the selfdual point $W=1$; see Figs.$~$\ref{fig:initial_proj}(c)-$~$\ref{fig:initial_proj}(e). However, with lower $W$, there are fewer and fewer initial states that fulfill the criterion in Eq.$~$\eqref{eqn:projection2}; see the yellow parts of the upper peaks in Figs.$~$\ref{fig:initial_proj}(c)-$~$\ref{fig:initial_proj}(e). The effect of the mobility edges should decrease with $W$ approaching the selfdual point $W=1$, and it should be minimal at the selfdual point.

For $W<1$, the roles of the real and dual momentum space switch. Here, for $1/W_c<W<1$, the mobility edges separate the states that are truly delocalized from those localized in the dual momentum lattice. Analogously, at $W=1/W_c\approx0.45$, the mobility edges separate the states that are truly delocalized from the critical states  in the dual momentum lattice. However, the transport of the initial localized state $|l_0\rangle$ becomes unable to distinguish the mobility edges in the dual momentum space, as it is sensitive only to the real space structure of the eigenstates. Lastly, for $W=0.1$, there are no mobility edges; all eigenstates are localized in momentum space and thus delocalized in real space. The initial localized state $|l_0\rangle$ projects roughly uniformly across the whole eigenspectra; see Fig.$~$\ref{fig:initial_proj}(e). For $W=0.1$, all $|l_0\rangle$ represent the same transport, which is expected to be close to ballistic.

To summarize, in order to avoid the effects of the spectral edges, we study initial states $|l_0\rangle$ that fulfill the criterion in Eq.$~$\eqref{eqn:projection2} for $W\geq1$. For $W<1$ we relax this criterion for $|l_0\rangle$. Reciprocally, for initial states $|k^{\rm D}_0\rangle$ we require an analogous criterion, i.e.,
\begin{align}
{\rm proj}_{50}^{k^{\rm D}_0}> 0.8,
\label{eqn:projection3}
\end{align}
to be fulfilled for $W\leq1$, and we relax this criterion for $W>1$.

\section{System-size fluctuations in steady-state transport}
 \label{3D case transport}

\begin{figure}[t!]
    \includegraphics[width=0.45\textwidth]{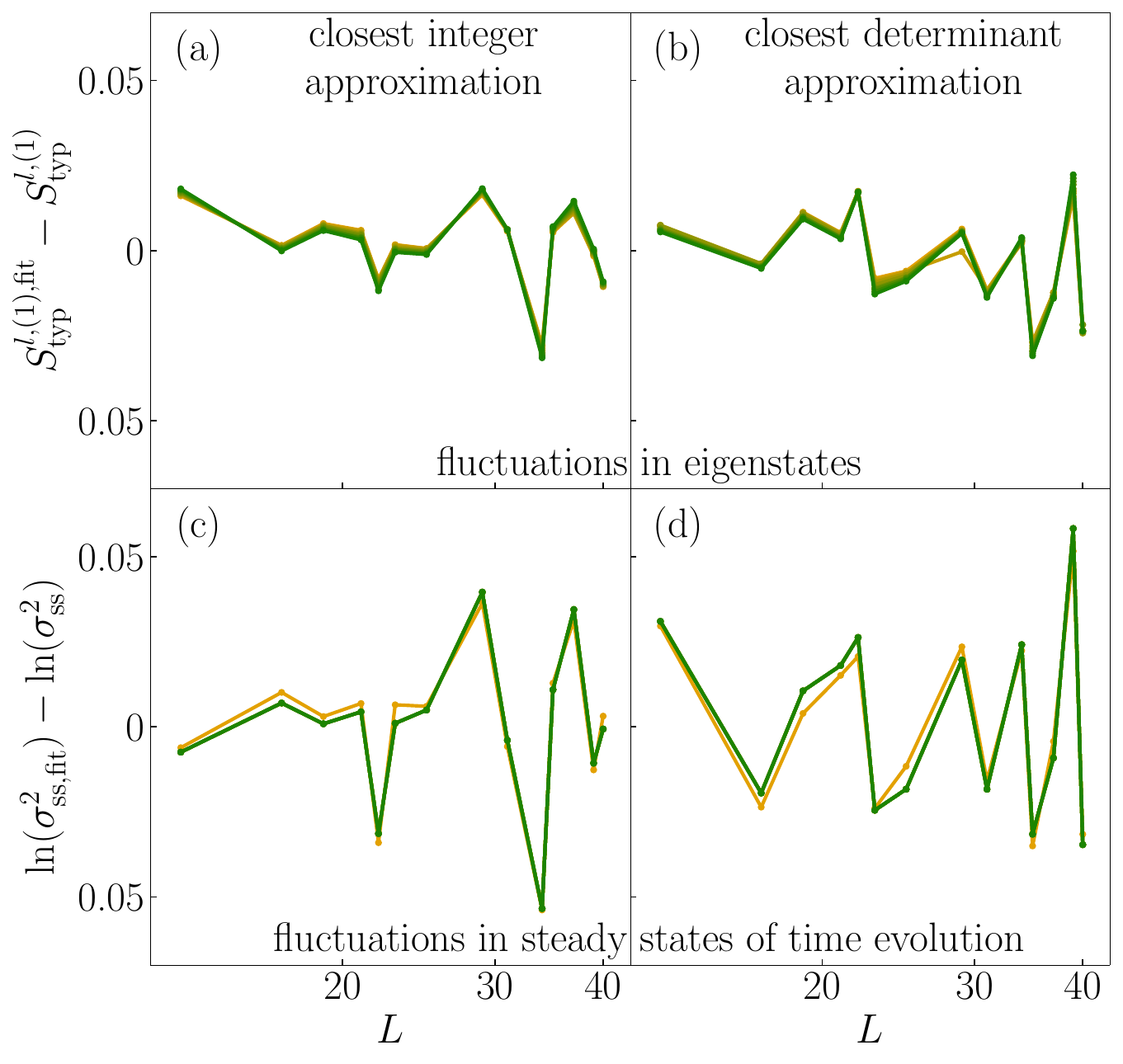} 
  \caption{Fluctuations of the wave-function and the steady-state transport measures  above the main trends in
  the 3D Aubry-André model for $\theta=\pi/7$ and $\beta=(\sqrt{5}-1)/2$ in (a), (c) the closest integer approximation of the $B$-matrix, and (b), (d) the closest determinant approximation of the $B$-matrix.  Panels (a)-(b) show the fluctuations of the participation entropy $S_{\rm typ}^{l,(1)}$ above linear fit $S_{\rm typ}^{l,(1),{\rm fit}}$ and panels
  (c) and (d) show logarithmic differences $\ln(\sigma^2_{\text{ss,\ fit}}) - \ln(\sigma^2_{\text{ss}})$ as a function of linear system size $L$, for $W=2.18, 2.19,...,2.25$ around the critical point $W_c$.}
  \label{fig:diff_sigma_ss_avg}
\end{figure}

To further support our claim that the transport reflects the physics of the eigenstates, we study the system-size fluctuations of $\sigma^2_{\text{ss}}$ above $\sigma^2_{\text{ss,\ fit}}$. The latter was discussed in Sec.$~$\ref{3D case time steady state}. In Fig.$~$\ref{fig:diff_sigma_ss_avg}, we compare logarithmic differences $\ln(\sigma^2_{\text{ss,\ fit}}) - \ln(\sigma^2_{\text{ss}})$, see Fig.$~$\ref{fig:diff_sigma_ss_avg}(c) and $~$\ref{fig:diff_sigma_ss_avg}(d), to the differences of the eigenstate participation entropies, see Figs.$~$\ref{fig:diff_sigma_ss_avg}(a) and$~$\ref{fig:diff_sigma_ss_avg}(b), which are discussed in Appendix$~$\ref{3D case fluctwave}. Remarkably, we indeed observe similar profiles of the fluctuations within each of the commensurate approximations.

\section{Spectral fractal dimension}
 \label{3D case d_s}

In this section, we show that the spectral fractal dimension $d_s$ has the trivial value $d_s=1$ for all $W$ values, as expected for simple spectra with the GUE, Poisson, or intermediate statistics. The spectral fractal dimension  can be extracted either from the scaling of the typical Heisenberg time with the system volume $V$ or from the box-counting method applied to the spectrum$~$\cite{hopjan2025critical}. 

\begin{figure}[b!]
    \includegraphics[width=0.49\textwidth]{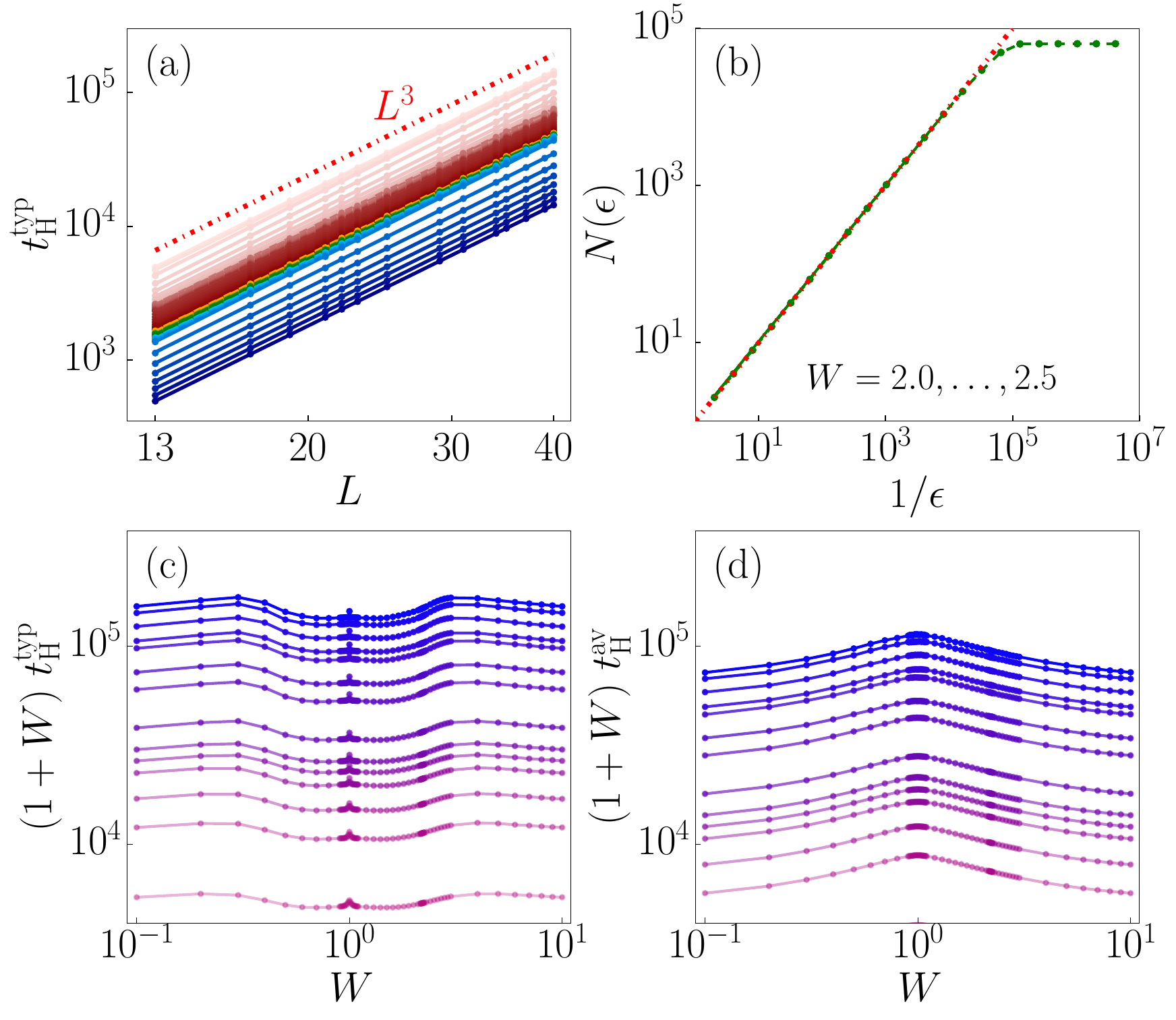} 
  \caption{Scaling of (a) the typical Heisenberg time, $t_{H}^{\rm{typ}}$, with linear systems size $L$, for various $W$ values
  (red to blue color), and (b) scaling of $N(\epsilon)$ with the inverse box length $1/\epsilon$, see Eq.$~$\eqref{N_scaling}, for various $W$ values close to $W_c$ in the 3D Aubry-André model, averaged over 100 Hamiltonian realization, $\theta=\pi/7$ and $\beta=(\sqrt{5}-1)/2$ in the closest integer approximation. (c) The symmetry of the rescaled the typical Heisenberg time, $t_{H}^{\rm{typ}}/(1+W)$, as a function of $W$. (d) The symmetry of the rescaled the average Heisenberg time, $t_{H}^{\rm{av}}/(1+W)$, as a function of $W$.}
  \label{fig:t_H_scaling_ds}
\end{figure}

The typical Heisenberg time is defined as the inverse of the typical eigenvalue spacing times $2\pi$,
\begin{equation}\label{Heis_typ_def}
    t_{H}^{\rm{typ}}= \frac{2\pi}{\delta E^{\rm{typ}}}=\frac{2\pi}{{\langle\exp(\langle {\ln({E_\mu-E_{\mu-1})}\rangle_\mu})\rangle_H}}\;,
\end{equation}
and it was shown$~$\cite{hopjan2025critical} that it scales as
\begin{equation}\label{Heis_typ_def_scling}
    t_{H}^{\rm{typ}}\propto V^{1/d_s} = L^{3/d_s}.
\end{equation}
We numerically test Eq.$~$\eqref{Heis_typ_def_scling} in Fig.$~$\ref{fig:t_H_scaling_ds}(a). We observe $L^{3}$ scaling for all values of $W$, implying that $d_s=1$ for all $W$. This confirms our expectation for the simple spectra with the GUE, Poisson, or intermediate statistics. In Fig.$~$\ref{fig:t_H_scaling_ds}(c), we further demonstrate the symmetry of $t_{H}^{\rm{typ}}$ when rescaled by $1/(1+W)$. We note that, as $d_s=1$, the averaged Heisenberg time,
\begin{equation}\label{Heis_av_def}
    t_{H}^{\rm{av}}= \frac{2\pi}{\delta E^{\rm{av}}}=\frac{2\pi}{{\langle\langle {{E_\mu-E_{\mu-1}}\rangle_\mu}\rangle_H}}\;,
\end{equation}
is proportional to the typical Heisenberg time, $t_{H}^{\rm{typ}}\propto t_{H}^{\rm{av}}$. In Fig.$~$\ref{fig:t_H_scaling_ds}(d), we demonstrate the symmetry of $t_{H}^{\rm{av}}$ when rescaled by $1/(1+W)$.

Finally, as a cross-check, we also extract the spectral fractal dimension using the box counting method. To this end, we transform the eigenenergies $E_\mu$ as $\epsilon_\mu=(E_\mu-E^{\rm min}_\mu)/(E^{\rm max}_\mu-E^{\rm min}_\mu)$, where $E^{\rm max}_\mu$ is the maximal energy and $E^{\rm min}_\mu$ is the minimal energy, such that the transformed eigenspectrum $\epsilon_\mu$ spans the interval $[0,1]$. The latter is then divided into $1/\epsilon$ boxes $B$ of length $\epsilon$. The box scaling function counts the number of non-empty boxes,
\begin{equation}\label{N_q_scaling}
      N_{}^{}(\epsilon) =\Big\langle \sum_{B\neq B_\varnothing} 1 \Big\rangle_H,
\end{equation}
and its scaling can be expressed as 
\begin{equation}\label{N_scaling}
     N(\epsilon) \propto \Bigl(\frac{1}{\epsilon}\Bigr)^{d_s}\;.
\end{equation}
We numerically test Eq.$~$\eqref{N_scaling} in Fig.$~$\ref{fig:t_H_scaling_ds} for $W$ close to the transition point. Again, we observe that $d_s=1$.

\bibliographystyle{biblev1}
\bibliography{references,references1,references2}

\end{document}